\documentclass[12pt,english,longbibliography,nofootinbib,superscriptaddress,12pt,sort&compress,showkeys]{extarticle}
\usepackage[T1]{fontenc}
\usepackage[latin9]{inputenc}
\usepackage{xcolor}
\usepackage{float}
\usepackage{booktabs}
\usepackage{multirow}
\usepackage{amsmath}
\usepackage{amssymb}
\usepackage{graphicx}
\usepackage{esint}
\usepackage[numbers]{natbib}
\PassOptionsToPackage{normalem}{ulem}
\usepackage{ulem}

\makeatletter

\providecommand{\tabularnewline}{\\}
\providecolor{lyxadded}{rgb}{0,0,1}
\providecolor{lyxdeleted}{rgb}{1,0,0}

\DeclareRobustCommand{\lyxsout}[1]{\ifx\\#1\else\sout{#1}\fi}

\usepackage{indentfirst}
\usepackage{amsfonts}
\usepackage[T1]{fontenc}
\usepackage{ae,aecompl}
\usepackage{sidecap}
\usepackage[section]{placeins}
\usepackage{epsf}
\makeatother

\usepackage{babel}
\date{\today}

\usepackage{chngcntr}

\makeatother

\usepackage{babel}
\begin{document}
\title{Atomistic modeling of metal-nonmetal interphase boundary diffusion\textbf{ }}
\author{I. Chesser$^{1}$, R. K. Koju$^{1}$, A. Vellore$^{2}$ and Y. Mishin$^{1}$\\
$^{1}${\normalsize{}Department of Physics and Astronomy, MSN 3F3},\\
 {\normalsize{}George Mason University, Fairfax, VA 22030, USA}\\
$^{2}${\normalsize{}Thomas Jefferson High School for Science and
Technology, Alexandria, VA 22312, USA }}

\maketitle
Atomistic computer simulations are applied to investigate the atomic
structure, thermal stability, and diffusion processes in Al-Si interphase
boundaries as a prototype of metal-ceramic interfaces in composite
materials. Some of the most stable orientation relationships between
the phases found in this work were previously observed in epitaxy
experiments. A non-equilibrium interface can transform to a more stable
state by a mechanism that we call interface-induced recrystallization.
Diffusion of both Al and Si atoms in stable Al-Si interfaces is surprisingly
slow compared with diffusion of both elements in Al grain boundaries
but can be accelerated in the presence of interface disconnections.
A qualitative explanation of the sluggish interphase boundary diffusion
is proposed. Atomic mechanisms of interphase boundary diffusion are
similar to those in metallic grain boundaries and are dominated by
correlated atomic rearrangements in the form of strings and rings
of collectively moving atoms. 

\date{}

\medskip{}


\section{Introduction\label{sec:Introduction}}

Metal-ceramic interfaces play an important role in many processes
in structural and electronic materials \citep{Balluffi95,Howe97,Zhu:2021aa}.
In particular, they often control the mechanical behavior of metal-matrix
composites reinforced with ceramic inclusions \citep{Wunderlich:2014aa,Zhang:2015ac,Damadam:2017ab,Yang:2018ad,Buehler:2019aa,Xu:2020aa,Lien:2020aa,Abboud:2020aa,Zhu:2021aa,Lien:2022aa}.
During the composite deformation at low and medium temperatures, the
damage usually starts at matrix-reinforcer interfaces and accumulates
by several concurrently operating mechanisms, including the interface
shearing, debonding, and interaction with dislocations coming from
the metallic matrix. During creep deformation, the interface resistance
to shear ensures that the reinforcer carries much of the applied load,
imparting a high creep resistance to the entire material \citep{Shen:1996aa,Ranganath96,Farghalli01,Huang:2003aa}.
However, at high enough temperatures, the interfaces lose much of
their shear resistance and the load shifts back to the metallic matrix.
As a result, the creep resistance decreases and can even fall below
that of the matrix material \citep{Rosler:1991aa}. 

The high-temperature sliding at metal-ceramic interfaces is not accompanied
by interface debonding or any visible accumulation of damage. This
process is fundamentally different from the low-temperature sliding
friction and has been identified as a distinct physical phenomenon
called interfacial creep \citep{Funn:1998aa,Dutta:2000aa,Peterson:2003aa}.
Although the microscopic mechanisms of interfacial creep remain largely
unknown, the process is clearly controlled by diffusive mass transport
of the metallic elements along the metal-ceramic interface. 

Much of the effort to understand the interfacial creep has focused
on the binary Al-Si system as a simple model of a composite material.
This system combines a soft metal with a hard covalent material representing
the ceramic phase. The system features a simple eutectic phase diagram
with low mutual solubility in the solid state, making it ideal for
a fundamental study. Peterson et al.~\citep{Peterson:2002aa,Peterson:2003aa,Peterson:2004aa}
studied interfacial creep at individual Al-Si interfaces using a single
crystal of Si sandwiched between two polycrystalline Al layers. The
interfacial creep rates measured at a series of loads and temperatures
gave the activation energy of 42 kJ/mol, which is smaller than the
experimental activation energy of Al grain boundary (GB) diffusion
(84 kJ/mol). However, this result could not be interpreted unambiguously
due to the highly non-equilibrium structure of the interfaces. High-resolution
transmission electron microscopy (TEM) revealed a 20 nm thick amorphous
interfacial layer stabilized by the oxygen coming from the native
oxide layer on the Al surface. Attempts to gain insights into the
diffusional creep at Al-Si interfaces were also made by other authors
\citep{Chen:2000ab,Dutta:2000aa,Dutta:2001aa}, but little progress
has been made in understanding the underlying mechanisms. 

A critical ingredient for studying the interfacial creep is the knowledge
of interface diffusion coefficients. Unfortunately, while diffusion
along metal-metal interphase boundaries has been measured in several
systems \citep{Kaur89,Kaur95,Sommer:1996aa,Minkwitz:1998aa}, little
is known about diffusion along metal-nonmetal interfaces. Qualitatively,
it is well-recognized that interface diffusion is faster than lattice
diffusion \citep{Kaur95}. However, there is no consensus regarding
the placement of metal-nonmetal interfaces relative to GB diffusion,
surface diffusion, and diffusion in liquids. Experimental measurements
of interphase boundary diffusion are extremely challenging. The only
direct measurement known to us was for indium chemical diffusion along
Sn-Ge twist interfaces \citep{Straumal:1984aa}. Among indirect measurements,
Kosinova et al.~\citep{Kosinova:2015aa} back-calculated Au diffusion
coefficients along Au-sapphire interfaces by comparing experimental
observations of solid-state dewetting with a kinetic model of the
process. The diffusion coefficients obtained were comparable with
those for Au diffusion along non-singular surfaces. Kumar et al.~\citep{kumar2018anomalous}
studied the kinetics of partial solid-state dewetting of a Ni film
from a sapphire substrate and estimated the Ni diffusivity along the
Ni-Al$_{2}$O$_{3}$ interface. Their results, supported by first-principles
density-functional theory (DFT) calculations, suggest that the rate
of Ni diffusion along this interface is close to the rate of Ni self-diffusion
in Ni GBs. On the other hand, Barda et al.~\citep{barda2020metal}
applied a similar experimental method and found that Au hetero-diffusion
along the same Ni-Al$_{2}$O$_{3}$ interface was slower than Au hetero-diffusion
in Ni GBs. 

Specifically for Al-Si interphase boundaries, we are unaware of any
direct or indirect diffusion measurements. The most accurate and direct
diffusion measurements are made with radioactive isotopes. Unfortunately,
Al does not have a suitable isotope for diffusion measurements \citep{Mehrer2007}.
The only suitable radioactive isotope of Si is $^{31}$Si, which has
a short lifetime (2.6 hours). This drastically limits the time for
diffusion experiments making them highly challenging \citep{Mehrer2007}.
Under the circumstances, atomistic computer simulations offer the
only viable option for studying the interface diffusion in Al-Si.
In addition to predicting diffusion coefficients, atomistic simulations
can help discover the atomic-level mechanisms of the diffusion process. 

Previous atomistic simulations of the Al-Si system were conducted
at room temperature or 0 K. They additionally assumed that the interfaces
were atomically sharp and separated pure Al from pure Si. It is well-established,
however, that interfaces are more diffuse at high temperatures than
at room temperature and separate solid solutions rather than pure
elements. Furthermore, the previous simulations primarily targeted
the cohesive strength and shear resistance of the interfaces, and
the interface interactions with dislocations coming from the metallic
side and absorbed by or transmitted through the interface \citep{Salehinia:2014aa,Damadam:2017aa,Damadam:2017ab,Rasheed:2017aa,Wang:2017aa,wu2022atomistic,Misra:2021aa}.
To our knowledge, no simulations have been performed for mass transport
along Al-Si interphase boundaries. The most relevant simulation work
has been published by Wu et al.~\citep{wu2022atomistic}, who calculated
Al vacancy formation and migration energies at the Al (111) $||$
Si (111) interface. The results suggest the possibility of faster
interface diffusion relative to Al self-diffusion in the lattice,
but no comparison with GB diffusion was discussed.

This paper reports on atomistic computer simulations of Al and Si
diffusion along Al-Si interphase boundaries at relatively high temperatures
up to the eutectic point. A significant part of this work was dedicated
to constructing such boundaries with diverse orientation relationships
between the phases and evaluating their structural stability during
high-temperature anneals. The interface diffusion coefficients were
extracted from direct molecular dynamics simulations without any adjustable
parameters or model assumptions other than the approximations underlying
the interatomic potential. To put our results in perspective, we also
computed the diffusion coefficients of both elements in GBs and the
bulk liquid phase. The collective nature of the interface diffusion
mechanisms was investigated and compared with that in Al GB diffusion.
In section \ref{sec:Discussion}, we propose a qualitative explanation
of the sluggish interphase boundary diffusion in the Al-Si system
found in this work.

\section{Methods\label{sec:methods}}

\noindent The Large-scale Atomic/Molecular Massively Parallel Simulator
(LAMMPS) \citep{LAMMPS} was utilized to conduct Monte Carlo (MC)
and molecular dynamics (MD) simulations. The software package OVITO
\citep{OVITO} was used to visualize and analyze interface structures
and diffusion mechanisms. A semi-empirical interatomic potential \citep{saidi_aeam}
was used to model interatomic bonding in the Al-Si system. This potential
was made compatible with 2022 versions of LAMMPS as part of the publicly
available lammps-plugins package \citep{plugins}.

\subsection{Phase diagram calculations}

Simulations of the Al-Si interphase boundaries require the knowledge
of the Al-Si phase diagram predicted by the interatomic potential.
Experimentally, the Al-Si system has a simple eutectic phase diagram
with a wide solid-solid miscibility gap and relatively small solubility
limits in the terminal solid solutions. The Al-based and Si-based
solutions have face-centered cubic (FCC) and diamond cubic (DC) crystalline
structures, respectively. They will be referred to as simply Al and
Si when no ambiguity can arise. Although the phase diagram predicted
by this potential was previously computed \citep{saidi_aeam} by a
thermodynamic integration technique, we decided to verify the diagram
by independent calculations using the phase coexistence method applied
in prior work \citep{mishin2004atomistic,NiAl,CuAg,NiCr}. The solidus
and liquidus lines on the phase diagram were computed by joining a
solid layer and a liquid layer with periodic boundary conditions,
so the system effectively contained two parallel solid-liquid interfaces.
The initial phase compositions were chosen according to the phase
diagram computed previously \citep{saidi_aeam}. The system was subjected
to a hybrid MC/MD anneal in which alternating blocks of MC swap attempts,
and MD integration steps were performed in the isothermal-isobaric
(NPT) ensemble at a set temperature. Swap attempts involved exchanges
of chemical types of randomly selected pairs of Al and Si atoms and
were accepted or rejected according to the Metropolis criterion. The
approach to equilibrium was accompanied by solid-liquid interface
motion and compositional changes in both phases. Equilibrium was achieved
when the interfaces ceased to migrate and continued to fluctuate around
constant positions while the bulk phase compositions reached steady-state
values. The solidus and liquidus compositions were then calculated
by averaging the compositions of bulk phase regions at least 2 nm
away from the interfaces. To verify the system size convergence, multiple
system sizes were tested both parallel and normal to the interface
plane at several temperatures. Examples of convergence tests can be
found in Supplementary Fig.~\ref{fig:figS2}. The results reported
below are for a system containing around 20,000 atoms, the interface
cross-section of 2 nm by 2 nm, and the initial liquid and solid phase
thicknesses of 20 nm each.

The solvus line calculations were performed using a similar solid-solid
phase coexistence method. The known disadvantage of this approach
is the low mobility of solid-solid interfaces. Nevertheless, convergence
was achieved at several temperatures close to the eutectic temperature
$T_{\mathrm{eu}}$ for a set of interfaces with sufficient mobility.
Such interfaces exhibited a step flow migration mechanism in which
Si atoms were added to existing Si steps at the interface. The interface
migration distances observed were 0.5 nm or less. The solid-solid
systems contained 5000 atoms with 6 nm thick Al and Si layers.

\subsection{Interphase boundary construction and equilibration\label{subsec:Traditional-construction}}

Two methods were applied to create equilibrium interphase boundaries:
the direct bonding method and the simulated epitaxy approach.

In the direct bonding method, unrelaxed interface structures were
created by bonding Al and Si grains of specified crystallographic
orientations along a planar interface with periodic boundary conditions
parallel to the interface plane ($X$ and $Y$ coordinate axes) and
free-surface boundary conditions normal to the interface ($Z$ axis).
Each grain had a thickness of at least 6 nm. Al and Si lattices exhibit
a significant misfit with the 0 K ratio of the lattice parameters
along the $\left\langle 100\right\rangle $ direction equal to $a_{\mathrm{Si}}/a_{\mathrm{Al}}=1.34$.
The interface plane dimensions were chosen sufficiently large to ensure
a 1\% or smaller misfit strain along the $X$ and $Y$ directions.
The misfit strain was applied homogeneously to the Al grain to meet
periodic boundary conditions. This misfit strain varied slightly with
temperature according to the precomputed thermal expansion coefficients
of Al and Si at the temperature of simulations. Several smaller systems
with larger misfit strains were studied in the context of the interface-induced
recrystallization effect described below.

The interface structure was relaxed by hybrid MC/MD simulations to
achieve structural and chemical equilibration. To this end, a fraction
of randomly selected Al atoms were converted to Si atoms. This fraction
was adjusted to match the estimated solid solubility of Si in Al along
the solvus line on the computed Al-Si phase diagram at the chosen
temperature. Next, an MC/MD anneal was performed with alternating
blocks of MC swap attempts and MD integration steps in the canonical
NVT ensemble. The number of swap attempts in each MC block was equal
to twice the number of Al atoms in the system. The MD blocks were
chosen to be 80 ps long, which was enough to sample atomic rearrangement
events distinct from vibrations. Throughout the annealing procedure,
a 1 nm slab containing the bottommost Si surface was frozen in place
and excluded from dynamics. A 1 nm slab was also fixed at the top
of the Al layer and excluded from dynamics (apart from the initial
composition seed). This slab was allowed to float freely as a rigid
body during the MD steps to relieve pressure normal to the interface
and accommodate possible spontaneous sliding events at the interface.
Convergence of the MC/MD calculations was monitored by tracking the
total potential energy of the system, the concentration of Si in the
Al lattice, and the total number of disordered atoms at the interface
that had neither FCC nor DC coordination. At high temperatures approaching
the eutectic temperature, convergence was typically achieved after
several ns of MD time and $10^{6}$ MC swap attempts. Production runs
were then performed for 20-40 ns and several million swap attempts.

The simulated epitaxy method was applied to find thermodynamically
preferred interface orientation relationships as a function of the
Si substrate orientation, temperature, and system size. The method
permits a discovery of interfaces that are difficult to find by the
direct bonding methodology, such as the interfaces with small lattice
rotations away from high symmetry orientations.

As the starting point of the method, unrelaxed interfaces were created
by the direct bonding method described above. The Al layer was then
melted and re-solidified on top of the Si layer by the following multi-step
procedure. First, the Al layer was separated from the Si layer by
creating a 1 nm thick vacuum layer at the interface. The Si surface
was then annealed for 0.1 ns at the set temperature, with all atoms
of the Al layer held fixed. This step allowed for possible reconstruction
on the Si surface. Next, the Al layer was quickly heated to 1200 K
and held for 0.1 ns, with the Si layer excluded from dynamics. Since
the melted Al layer was allowed to float in the normal direction,
it closed the gap to wet the Si surface. The entire system was then
quenched to a target temperature and annealed in the NVT ensemble
for several ns using MD only. During the anneal, the Al layer solidified
epitaxially on the Si surface. The epitaxial orientation relationship
obtained was quantified using the grain segmentation and polyhedral
template matching (PTM) modifiers in OVITO \citep{PTM}. Selected
epitaxial structures were subsequently annealed with an additional
MC/MD run to examine the interface structure and verify the interface
stability.

\subsection{Grain boundary construction and equilibration}

A set of Al and Si GBs was created to investigate GB diffusion. Unrelaxed
single-component Al and Si GBs were constructed by joining two grains
in a simulation box with periodic boundary conditions in the GB plane
and free-surface boundary conditions normal to the GB. Supplementary
Table \ref{tab:S1} specifies the crystallography of each GB constructed.
The GB cross-section had dimensions of at least 10 nm by 10 nm, and
each grain had a thickness of at least 6 nm. A typical simulation
box contained 50,000 to 100,000 atoms. For commensurate GBs, the lateral
bicrystal dimensions were chosen with integer repeats of the coincident
site lattice unit cell. For the incommensurate GB with the $\left\langle 100\right\rangle $
and $\left\langle 110\right\rangle $ directions parallel to the $X$
axis, a small strain of approximately 1.001 was applied to the upper
grain to satisfy periodic boundary conditions.

The initial GB structures were optimized by a grid search that seeks
to find a deep energy minimum. Several hundred initial unrelaxed GB
structures were prepared with different rigid-body shifts applied
to the upper grain relative to the lower. For each initial structure,
pairs of closely spaced atoms were identified, and one atom was deleted
at random if the overlap radius was in the range (0.7-0.95)$r_{0}$,
where $r_{0}$ is the first nearest neighbor distance in FCC Al or
DC Si. Conjugate-gradient energy minimization was then performed at
0 K to relax the GB core atoms in each candidate structure. The lowest-energy
structure was used as input for the subsequent diffusion simulations.
This optimization procedure recovered the well-known structures of
the $\Sigma3$ and $\Sigma17$ GBs in Al and the $\Sigma17$ GB in
Si previously found with a Tersoff-type interatomic potential \citep{hickman2020thermal}.

The lowest-energy GB structures were further equilibrated by a multi-step
MD simulation in the NVT ensemble. First, the bicrystal was homogeneously
expanded by the precomputed lattice thermal expansion strain at the
chosen temperature. A 1 nm thick rigid slab at the top of the upper
grain was allowed to float in the normal direction during a short
(0.1 ns) NVT anneal. Combined with the pre-expansion, this step reduced
the stresses in the grains to near zero. Next, the rigid-body slabs
at the top and bottom of the bicrystal were fixed and an NVT anneal
was performed for up to 20 ns. The fixed boundary conditions suppressed
any spontaneous GB sliding events.

In addition to the elemental GBs, the $\Sigma3$ and $\Sigma17$ GB
structures were created in Al-Si alloys with a bulk concentration
of 8 at.$\%$ Si. The hybrid MC/MD method was employed for chemical
and structural equilibration using the protocol described previously
for the interphase boundaries. Convergence of Si GB segregation was
achieved within several ns of MD time and approximately $10^{6}$
MC swap attempts. Note that the alloy GBs represented a single-phase
state and thus had an additional degree of freedom compared with the
interphase boundaries. An alloy GB could be studied at any chemical
composition and temperature within the Al-rich single-phase domain
on the phase diagram. By contrast, only temperature could be varied
for a interphase boundary while the phase compositions were uniquely
defined by the solvus lines on the phase diagram.

\subsection{Interface diffusion coefficient calculations}

GB and interphase boundary self-diffusion coefficients were measured
by tracking the motion Al and Si atoms within a 2 nm thick layer centered
at the interface during the production MD runs in the microcanonical
(NVE) ensemble which lasted between 1 ns and 40 ns. The interface
position was determined as the average $Z$-coordinate of non-FCC
and non-DC atoms identified by the PTM modifier in OVITO \citep{PTM}.
The diffusion coefficients in the interface layer were calculated
from the Einstein relations $D_{x}=\langle x^{2}\rangle/2t$ and $D_{y}=\langle y^{2}\rangle/2t$,
where $\langle x^{2}\rangle$ and $\langle y^{2}\rangle$ are the
mean squared displacements (MSDs) of atoms in the in-plane $X$ and
$Y$ directions and $t$ is the simulation time. Bootstrap resampling
was employed as in \citep{race2015quantifying} to compute error estimates
associated with the diffusion coefficients. The hyper-parameters chosen
for this method included a smoothing window of 5 ps, a block length
of 20 ps, and a number of resampled trajectories equal to 100. To
mitigate the effect of lattice diffusion on the extracted interface
diffusion coefficients, only atoms that remained in the calculation
box at initial and final times were considered. The diffusion coefficients
were additionally averaged over at least three independent runs with
different initial composition and/or velocity seeds. For the interface
containing disconnections, the diffusion coefficients were averaged
over five initial velocity seeds.

Using a fixed-width layer underestimates the actual interface diffusion
coefficient because of the inclusion of immobile atoms from the adjacent
perfect lattice regions. For Al diffusion, this underestimation was
corrected by rescaling the diffusion coefficients by the inverse fraction
of perfectly coordinated FCC atoms averaged over the duration of the
simulation. For Si diffusion, the diffusion coefficients were rescaled
by the time-averaged inverse fraction of perfectly coordinated FCC
and DC atoms. The latter rescaling relies on the observation that
Si atoms occupy substitutional lattice positions in Al-Si alloys and
are FCC-coordinated. Local coordination was determined with the PTM
modifier \citep{PTM} in OVITO with a cutoff value of 0.4.

\subsection{Liquid diffusion coefficient calculations}

To compute liquid diffusion coefficients, an initial 32,000-atom liquid
structure was created in a periodic box by heating an Al crystal in
the NPT ensemble to 1500 K and holding it for 200 ps. Next, a set
of liquid samples was generated in the temperature range from 1050
to 1450 K by quenching the high-temperature structure at a rate of
50 K/100 ps and holding it at a set temperature for 200 ps. A further
stepwise quenching was performed from 1050 K to 600 K at a rate of
25 K/ns with a hold time of 2 ns every 50 K. These times were sufficient
for liquid equilibration at the chosen temperatures as evidenced by
converged total potential energy and linear MSD vs time behavior.

An equivalent procedure was used for Al-Si eutectic liquid with 27
at.$\%$ Si and pure Si liquid. For pure Si, a larger initial melting
temperature of 2000 K was employed. After obtaining well-equilibrated
structures, production NVE anneals were performed for all liquid structures
for 0.1 ns. The diffusivity was computed from the 3D Einstein relation
$D=\langle r^{2}\rangle/6t$ and averaged over five independent runs
with different velocity seeds. Crystallization was observed at sufficiently
low temperatures. Diffusion data for such cases is not reported here.

\subsection{Analysis of collective diffusion mechanisms}

The following algorithm was used to identify correlated string-like
clusters of mobile atoms during interface diffusion. First, mobile
atoms were identified within the interphase boundary or GB such that
the net displacement of an atom $\Delta r$ during a time interval
$\Delta t$ was within the range $0.4r_{0}<\Delta r<1.2r_{0}$. Here,
the upper bound was chosen to eliminate atoms that had undergone multiple
hops, and the lower bound was chosen to eliminate immobile atoms.
Next, mobile atomic pairs $(i,j)$ were found that remained nearest
neighbors at the times $t=0$ and $t=\Delta t$ and satisfied the
criterion $\text{min}(|\mathbf{r}_{i}(t)-\mathbf{r}_{j}(0)|,|\mathbf{r}_{j}(t)-\mathbf{r}_{i}(0)|)<0.43r_{0}$,
where $\mathbf{r}_{i}(t)$ is the dynamic trajectory of atom $i$.
This criterion identifies atomic pairs with string-like motion in
which one atom jumps into the previous position of the other. The
algorithm has three hyper-parameters: the lower and upper bounds for
the displacement and the substitution distance (the factors $0.4$,
$1.2$ and $0.43$ above). These parameters were chosen based on the
prior studies of diffusion in glass-forming supercooled liquids \citep{gbmdhpnas,AlSm1}
and were fixed across all interfaces studied in this work.

\section{Results}

\subsection{Al-Si phase diagram}

The Al-Si binary phase diagram computed in this work is shown in Supplementary
Fig.~\ref{fig:figS1}. It is topologically similar to the experimental
diagram \citep{EXPT_phase_dia} and differs quantitatively from the
diagram computed by Saidi et al.~\citep{saidi_aeam} with the same
interatomic potential. In particular, our calculations give a lower
eutectic temperature $T_{\mathrm{eu}}$ and a higher Si concentration
in the eutectic liquid than reported in \citep{saidi_aeam}. We believe
that our calculations are more accurate. The phase coexistence method
implemented in this work is direct and more reliable than the thermodynamic
integration method used in \citep{saidi_aeam}. To further validate
our calculations, we performed additional simulations involving phase
changes. For example, we observed melting of solid-solid systems when
they were heated above the eutectic temperature computed in this work
but below $T_{\mathrm{eu}}$ reported in \citep{saidi_aeam}.

The phase diagram predicted by this potential overestimates the maximum
solid solubility of Si in Al (0.12 $\pm$ 0.01 Si mole fraction) compared
with experiment (0.015) \citep{EXPT_phase_dia}. It also overestimates
the experimental eutectic composition (0.27 $\pm$ 0.02 versus 0.122).
Furthermore, although the Al and Si melting temperatures are reproduced
fairly accurately, the computed eutectic temperature (675 $\pm$ 5
K) is lower than the experimental value (850 K) \citep{EXPT_phase_dia}.
Other interatomic potentials for the Al-Si system \citep{dongare2012angular,starikov2020optimized}
report even lower $T_{\mathrm{eu}}$ values. Thus, while the interatomic
potential used in this work captures the correct topology of the Al-Si
phase diagram and has certain advantages over other potentials, it
does not reproduce the phase diagram with quantitative accuracy. Due
to this limitation, the Al-Si simulations reported in this paper are
not expected to make quantitatively accurate predictions for this
system. Rather, our goal is to explore general interface diffusion
trends in a model metal-nonmetal system featuring a simple eutectic
phase diagram.

\subsection{Interphase boundary structures}

Over a dozen crystallographically distinct Al-Si interphase boundaries
were studied in this work. They were created by either the direct
bonding method or by simulated epitaxy and equilibrated as described
in section \ref{subsec:Traditional-construction}.

\begin{figure}[H]
\centering\leavevmode \includegraphics[width=1\textwidth]{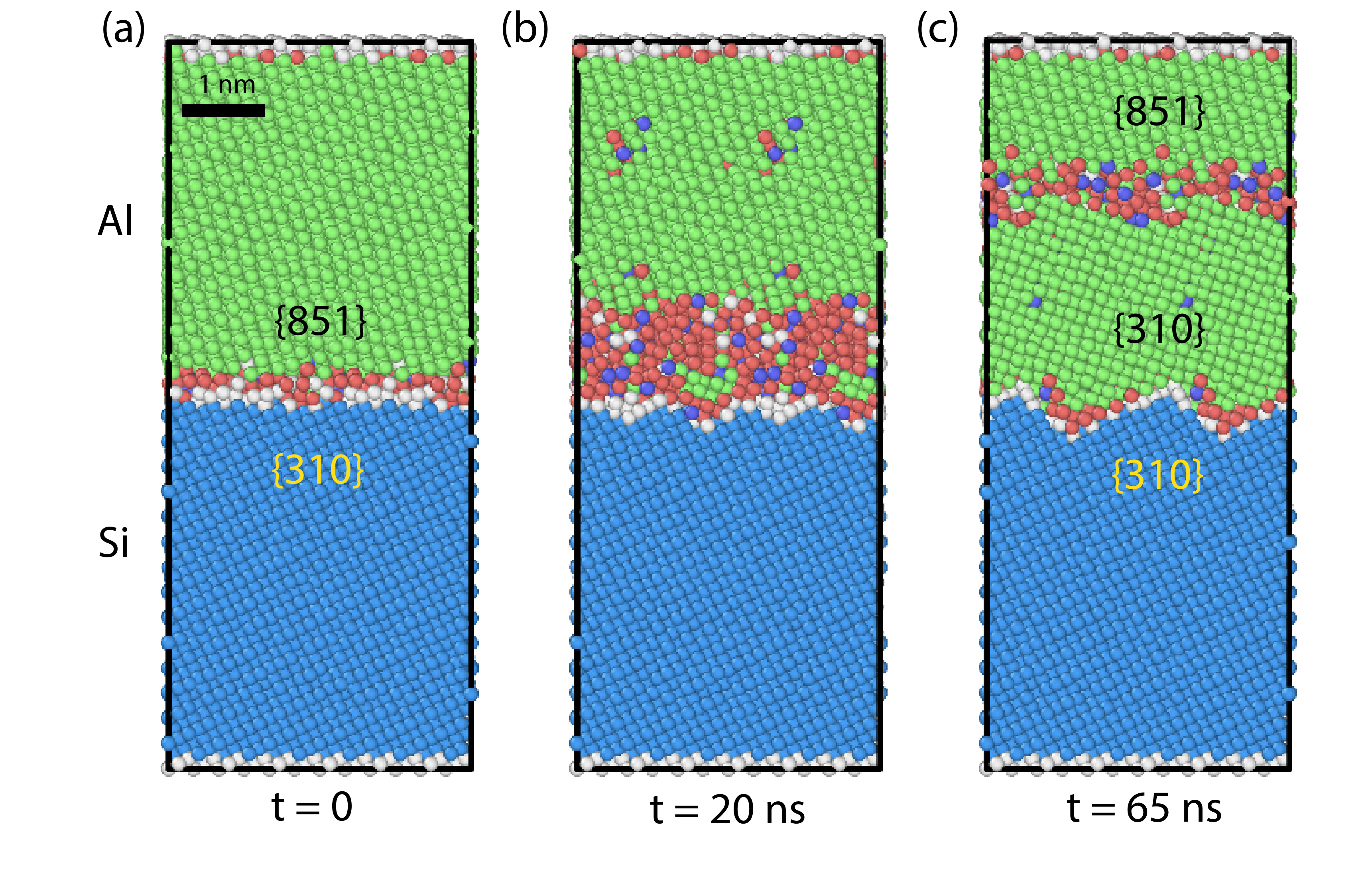}
\caption{Example of interface-induced recrystallization in the Al-Si system
at 650 K (0.96$T_{\text{eu}}$). MD time is indicated and atoms are
colored by local coordination: green = FCC, blue = DC, and all other
colors represent atoms in locally disordered environments. (a) Initial
Al-Si interface at early stages of equilibration. (b) New Al grain
nucleates at the interface. (c) Si-enriched GB migrates upward into
Al, leaving a faceted Al-Si interface behind. The interface orientation
relationships before and after recrystallization are indicated.}
\label{fig:fig1}
\end{figure}

A first key finding is that many interfaces with general crystallography
recrystallize into more stable crystallographic orientation relationships
at high temperatures in a process that we call interface-induced recrystallization.
An example is shown in Fig.~\ref{fig:fig1} for the $\left\{ 851\right\} _{\mathrm{Al}}||\left\{ 310\right\} _{\mathrm{Si}}$
interface. Upon annealing the initial interface for several ns, a
new Al grain nucleates at the interface with a cube orientation relationship
$\left\{ 310\right\} _{\mathrm{Al}}||\left\{ 310\right\} _{\mathrm{Si}}$.
Growth of the new grain occurs by rapid migration of a newly formed
Si-enriched GB into the parent Al grain. Si concentration at the new
GB is around 26 at.\%, which is close to the eutectic liquid composition
(27 at.\%Si). The interphase boundary left behind the moving GB is
strongly faceted and atomically sharp. Other general interphase boundaries
were also observed to recrystallize into thin, faceted structures
with near-cube orientation relationships (see Table \ref{tab:T1}).
Interesting additional features of the interface-induced recrystallization
phenomenon, such as solute drag during the GB migration and the misfit
strain dependence of this process, are discussed in the Supplementary
Information file.

\begin{figure}[H]
\centering\leavevmode \includegraphics[width=0.5\textwidth]{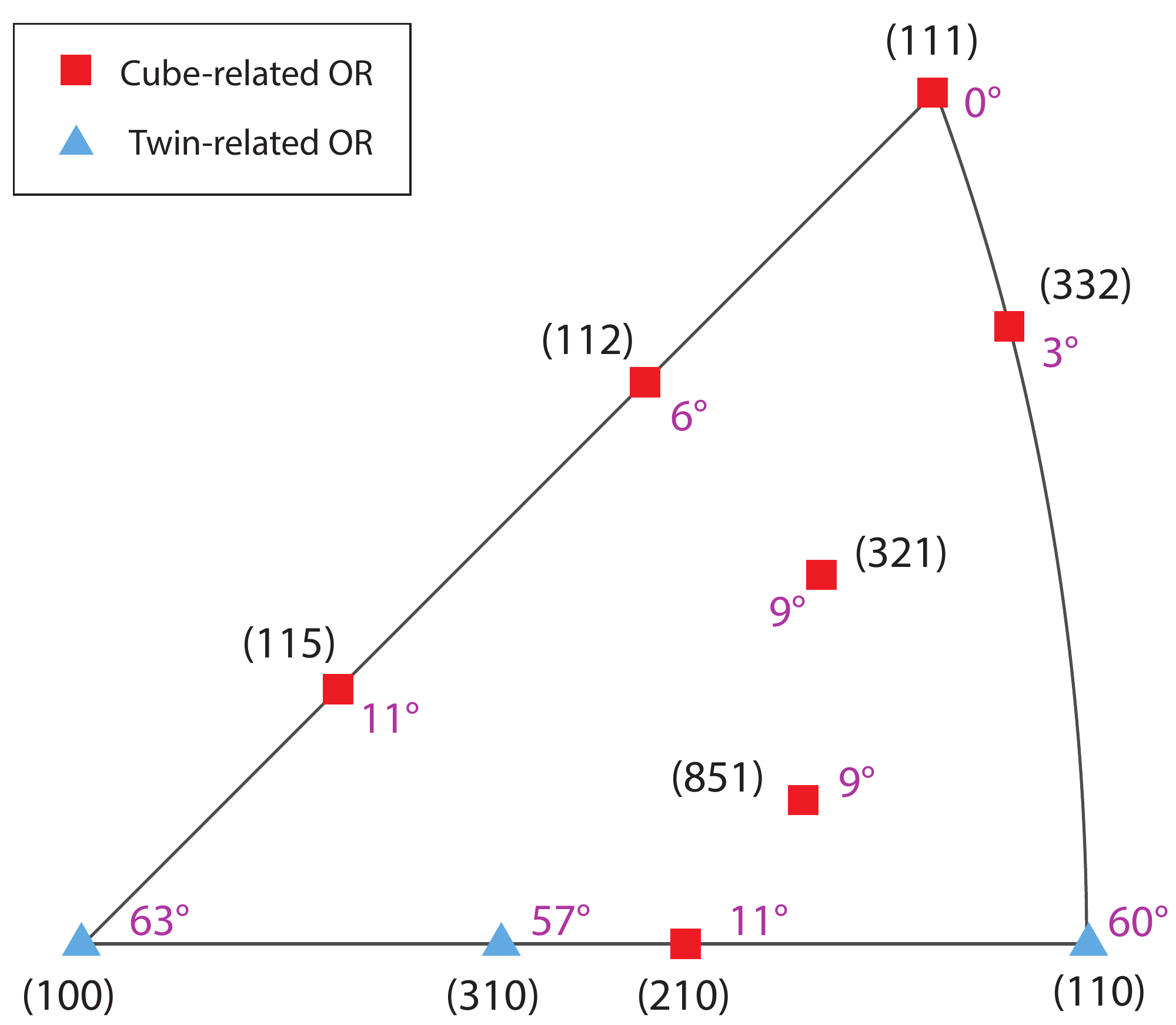}
\caption{Crystallographic orientation relationships at Al-Si interfaces produced
by simulated epitaxy. Si substrate orientations are indexed and plotted
in the stereographic triangle. The disorientation angles (shown in
purple) indicate deviations from the exact cube or twin orientations.}
\label{fig:fig2}
\end{figure}

A second key finding is that only a small set of interfaces with high-symmetry
orientation relationships were stable at high temperatures. We obtained
a set of stable interfaces via multiple processing methods, including
direct bonding, recrystallization, and simulated epitaxy. Table \ref{tab:T1}
summarizes all stable interfaces found in this work. The stable orientation
relationships produced by epitaxy simulations are graphically summarized
in Fig.~\ref{fig:fig2} for a range of Si substrate orientations.
The stable orientation relationships are either cube-type or twin-related
and are in good agreement with epitaxy experiments \citep{westmacott_review}.
For example, both the $\left\{ 111\right\} _{\mathrm{Al}}||\left\{ 111\right\} _{\mathrm{Si}}$
and $\left\{ 110\right\} _{\mathrm{Al}}||\left\{ 001\right\} _{\mathrm{Si}}$
interfaces remained stable in our simulations at high temperatures
and have also been found predominant over a range of temperatures
in epitaxy experiments \citep{westmacott_review,dahmen_100,meta_111}.
Recent simulation studies of near-eutectic Al-Si alloys also suggest
a predominance of cube orientation relationships \citep{wu2022atomistic},
which is consistent with our results.

An intriguing feature of the interfaces produced by recrystallization
and epitaxy is the lattice rotations away from the exact cube or twin-related
orientations. The magnitudes of such rotations vary from 0 to 11 degrees
and disagree across the processing methods. For example, epitaxially
deposited interfaces exhibit larger rotations away from the cube orientation
and have less pronounced faceted structures than recrystallized interfaces.
The rotations away from the exact cube or twin-related orientations
are reminiscent of the lattice rotations observed in epitaxy studies
of FCC-FCC metallic systems \citep{wynblatt2015importance,chatain2015importance,wynblatt2022heteroepitaxy}.
The structures constructed by direct bonding are constrained to have
zero disorientation. They also tend to have less pronounced faceting
than recrystallized structures. These observations demonstrate a rich
multiplicity of possible metastable variants of the cube and twin-related
interphase boundaries. 

\begin{figure}[H]
\centering\leavevmode \includegraphics[width=1\textwidth]{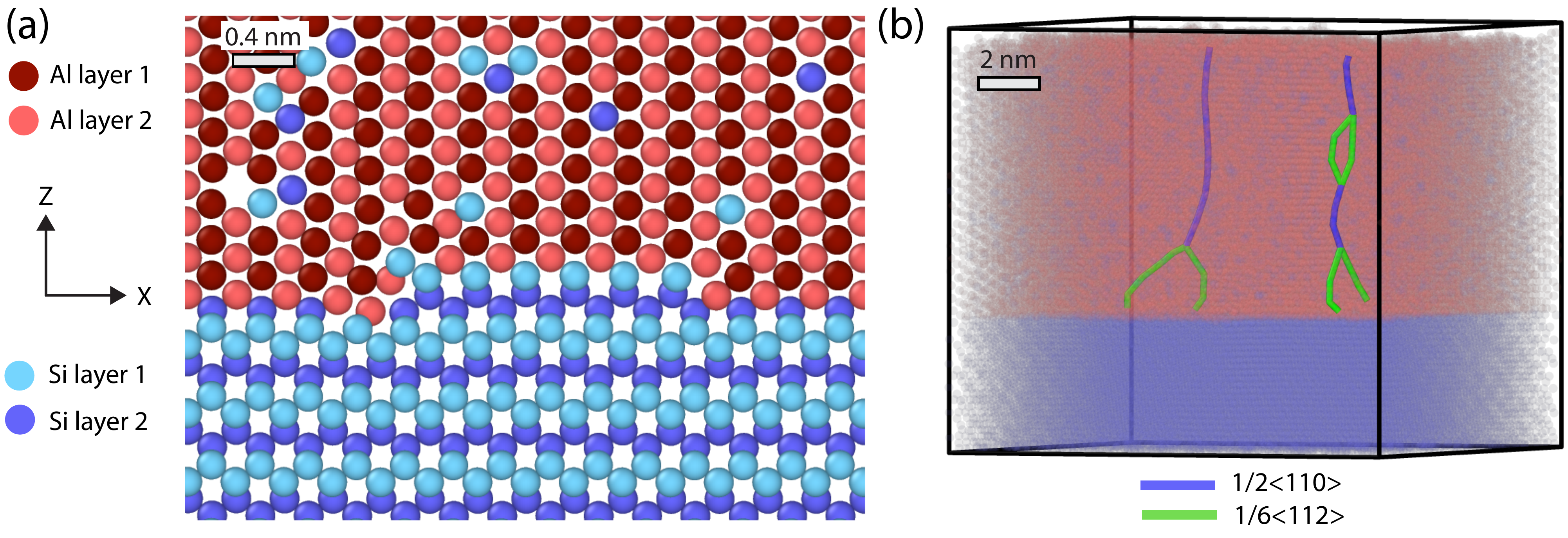}
\caption{Examples of defects at Al-Si interfaces. The interface orientation
is $\left\{ 110\right\} _{\mathrm{Al}}||\left\{ 001\right\} _{\mathrm{Si}}$,
with $\left\langle 100\right\rangle _{\mathrm{Al}}||\left\langle 110\right\rangle _{\mathrm{Si}}$
along the $X$ axis and $\left\langle 110\right\rangle _{\mathrm{Al}}||\left\langle 110\right\rangle _{\mathrm{Si}}$
along the $Y$ axis. The dark and light colors differentiate between
atomic positions in alternating layers parallel to the page ($X$-$Z$
plane). (a) Disconnection dipole in the directly bonded interface
with 13:14 ratio of Al:Si periods in the $X$ direction. The disconnection
Burgers vector is 0.194 nm parallel to $X$ (see Supplementary Fig.~\ref{fig:figS5}).
(b) The same interface produced by simulated epitaxy with a larger
(10 nm by 10 nm) cross-section does not contain disconnections but
creates threading screw dislocations in the Al grain, which dissociate
into partials at the interface. Both structures were observed at 650
K (0.96 $T_{\text{eu}}$).}
\label{fig:fig3}
\end{figure}

Deviations from perfect interface equilibrium were also revealed in
the form of additional defects in some of the interfaces, such as
disconnections and threading dislocations (Fig.~\ref{fig:fig3}).
These additional defects remained stable on the MD timescales accessible
to us (up to 100 ns) and served as additional ``short-circuit'' pathways
for diffusion. An example of interface disconnections is shown in
Fig.~\ref{fig:fig3}a for the $\left\{ 110\right\} _{\mathrm{Al}}||\left\{ 001\right\} _{\mathrm{Si}}$
interface constructed by the direct bonding method. Crystallographic
characterization of the disconnections is given in the Supplementary
Information file (Supplementary Fig.~\ref{fig:figS5}). The same
interface produced by simulated epitaxy exhibits threading dislocations
with perfect screw $1/2\left\langle 110\right\rangle $ character,
which dissociate into Shockley partials near the interface. These
examples demonstrate various ways by which the Burgers vector content
of an interphase boundary can be redistributed and localized. 

Melting behavior of the interfaces was examined at a range of temperatures
just below and above the eutectic temperature. It was found that,
in all cases, superheating was required (often by 100 K or more) to
cause melting on the MD timescales. Melting was observed to nucleate
at interface defects such as disconnections or facet junctions. In
previous work, two classes of interfaces were distinguished based
on whether pre-melting was observed before the bulk melting temperature
(disordered interfaces) or upon superheating (ordered interfaces)
\citep{tang2006diffuse}. The absence of pre-melting indicates that
the Al-Si interphase boundaries studied in this work are well-ordered
interfaces similar to coherent GBs. 

\subsection{Diffusion coefficients and mechanisms}

Diffusion in stable Al-Si interfaces was found to be very slow on
the MD timescale. Although some atomic hops were observed in the interface
region during long anneals, this atomic motion was insufficient for
reaching the diffusive regime in which the Einstein relation would
hold. Facet junctions at nano-faceted interfaces did not exhibit sufficient
atomic mobility either. The lower limit of diffusion coefficients
that could be measured in this work is about $D=10^{-13}$ m$^{2}$/s.
This number provides an upper estimate of possible Al-Si interface
diffusion coefficients. 

\begin{figure}[H]
\centering\leavevmode \includegraphics[width=1\textwidth]{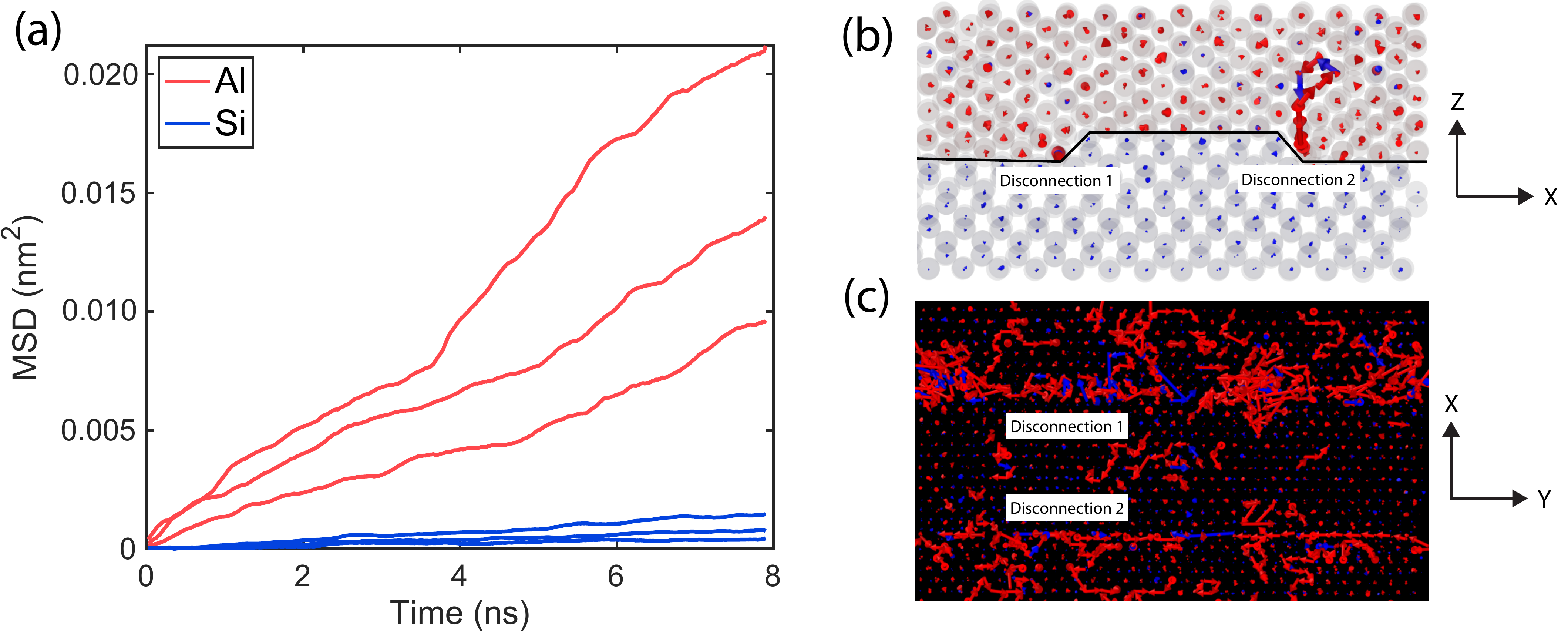}
\caption{Calculation of diffusion coefficients in the Al-Si interphase boundary\textbf{
}containing a disconnection dipole. (a) Example of MSD versus time
curves for diffusive displacements of Al and Si atoms parallel to
the disconnection lines ($Y$-axis). The three curves for each chemical
element correspond to the temperatures of 630 K, 640 K and 650 K in
the order of increasing diffusion coefficients. (b) Disconnection
diffusion is accompanied by occasional injection of displacement cascades
into the Al lattice which often return to the disconnection core (1
ns time interval shown). (c) The net displacement field in the interface
plane over the time interval of 8 ns indicates fast diffusion along
disconnection lines. In (b) and (c), atomic displacements are colored
by atomic type with Si displacements shown in blue and Al displacements
in red.}
\label{fig:fig4}
\end{figure}

The diffusive regime was successfully reached only for the $\left\{ 110\right\} _{\mathrm{Al}}||\left\{ 001\right\} _{\mathrm{Si}}$
interface containing a disconnection dipole. The interface diffusion
coefficients were extracted from the slopes of MSD versus time curves
illustrated in Fig.~\ref{fig:fig4}a. The curves display some degree
of diffusion intermittency manifested by the slight wiggles. However,
this intermittency is much weaker than the one observed for Al GB
diffusion \citep{Chesser:2022aa}, which was caused by point-defect
avalanches in the GB core. Collective diffusion mechanisms in the
disconnection core were visually apparent from atomic displacement
fields revealing string-like displacements of both Al and Si atoms
parallel to the disconnection lines (Fig.~\ref{fig:fig4}c). Disconnection
diffusion occasionally produced displacement cascades in nearby lattice
regions, which took winding paths through the Al grain before returning
into the disconnection (Fig. \ref{fig:fig4}b). Similar excursions
of point defects into the lattice were previously observed in MD simulations
of dislocation diffusion in Al \citep{Pun09a}. In this work, such
excursions were also found during Al GB diffusion, although they were
not as frequent as for disconnection diffusion. 

\begin{figure}[H]
\includegraphics[width=1\textwidth]{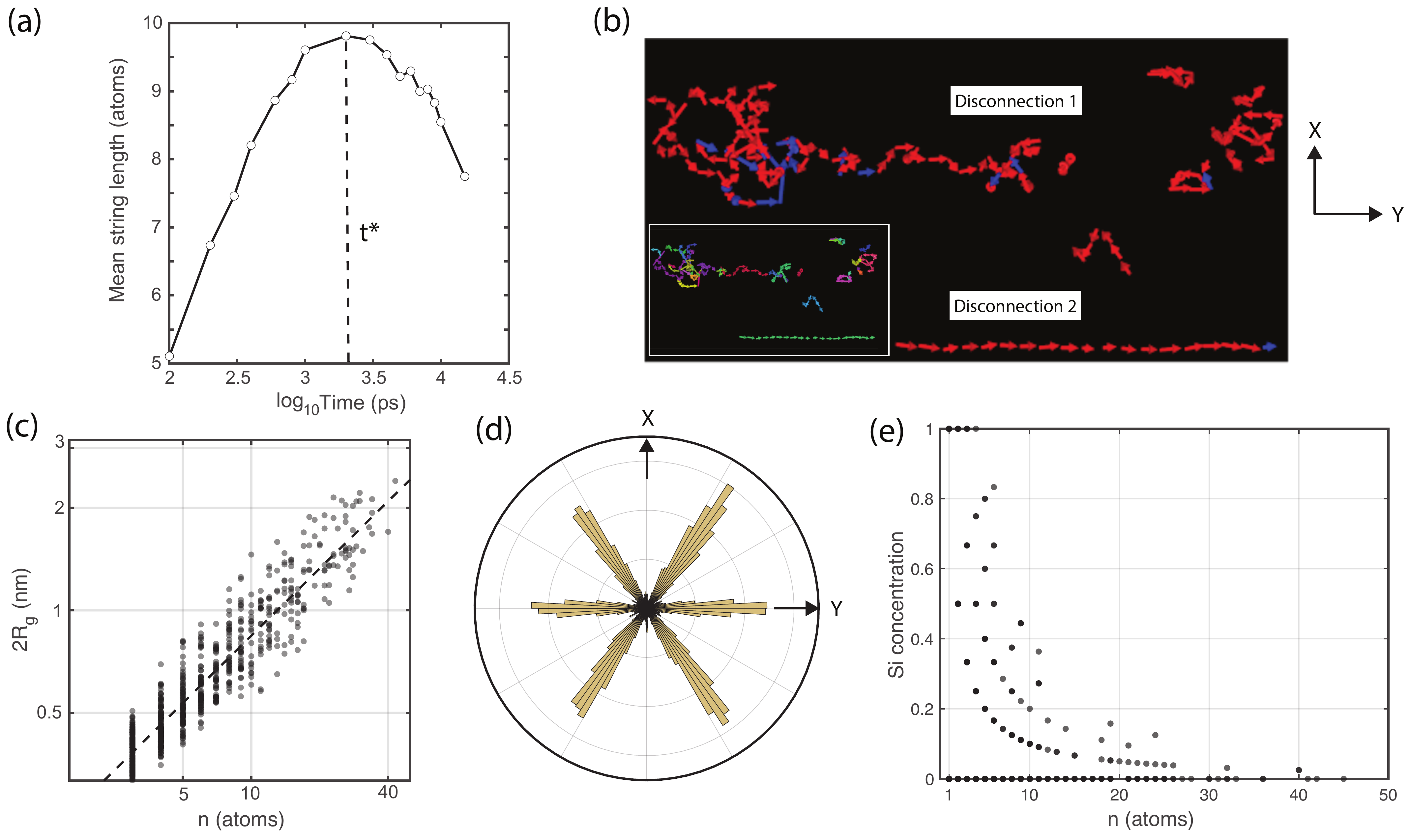} \caption{Collective diffusion mechanisms in Al-Si interphase boundaries. Diffusion
along the interphase boundary disconnections exhibits hallmarks of
dynamic heterogeneity, such as: (a) Characteristic timescale $t^{*}$
associated with the formation of mobile string-like atomic clusters.
The mean string contains about 9 atoms for top 50\% of mobile atoms.
(b) Examples of strings for the time interval of $t^{*}$ with string-like
displacements colored by atom type (blue = Si, red = Al). The string-like
clustering is shown in the inset with different colors representing
different clusters. (c) The fractal dimension of $d=1.5$ of string-like
clusters was extracted from the linear fit (dashed line) of the gyration
radius $R_{g}$ versus string size $n$ in logarithmic coordinates.
(d) Stereographic histogram of atomic displacement directions reveals
a propensity for the motion in the close-packed $\left\langle 110\right\rangle $
directions. (e) Si concentration (atomic fraction) in the strings
plotted as a function of string size. Si atoms play a lesser role
in the collective diffusion mechanisms than Al atoms but still participate
in the collective diffusion events.}
\label{fig:fig5}
\end{figure}

Disconnection diffusion mechanisms exhibit the familiar hallmarks
of dynamical heterogeneity, including spontaneous formation and dissolution
of string-like clusters of mobile atoms (Fig.~\ref{fig:fig5}a-c).
Examples of string-like displacement clusters are shown in Fig. \ref{fig:fig5}b-c
for the time interval $t^{*}$ at which the average string length
attains a maximum (Fig.~\ref{fig:fig5}a). The string-like clusters
have a fractal dimension of 1.5, which is halfway between a linear
chain and a random walk in 2D. A similar fractal dimension was found
in previous simulations of GB diffusion in FCC Cu \citep{mishin2015atomistic}.
Individual atomic displacements tend to have components along $\left\langle 110\right\rangle $
directions, as shown in the displacement orientation probability distribution
in Fig.~\ref{fig:fig5}e. Although Al atoms are the dominant participants
in the string-like motion, Si atoms are also involved in the strings.
For example, the linear chain of Al atomic displacements shown at
the top of Fig.~\ref{fig:fig5}c stops at a Si atom, which only hops
a short distance in the same direction. Si atoms also participate
in ring-like diffusion mechanisms and other collective events. Pure
Si strings of up to 4 atoms were observed, while some Al strings comprised
over 40 atoms (Fig.~\ref{fig:fig5}f). This contrast in the degree
of participation suggests that Si diffusion is associated with a smaller
length scale of collective motion compared with Al diffusion. We hypothesize
that Si atoms interrupt the collective mechanisms that would otherwise
have an even larger scale in Al GB disconnections.

\begin{figure}[H]
\centering\leavevmode \includegraphics[width=1\textwidth]{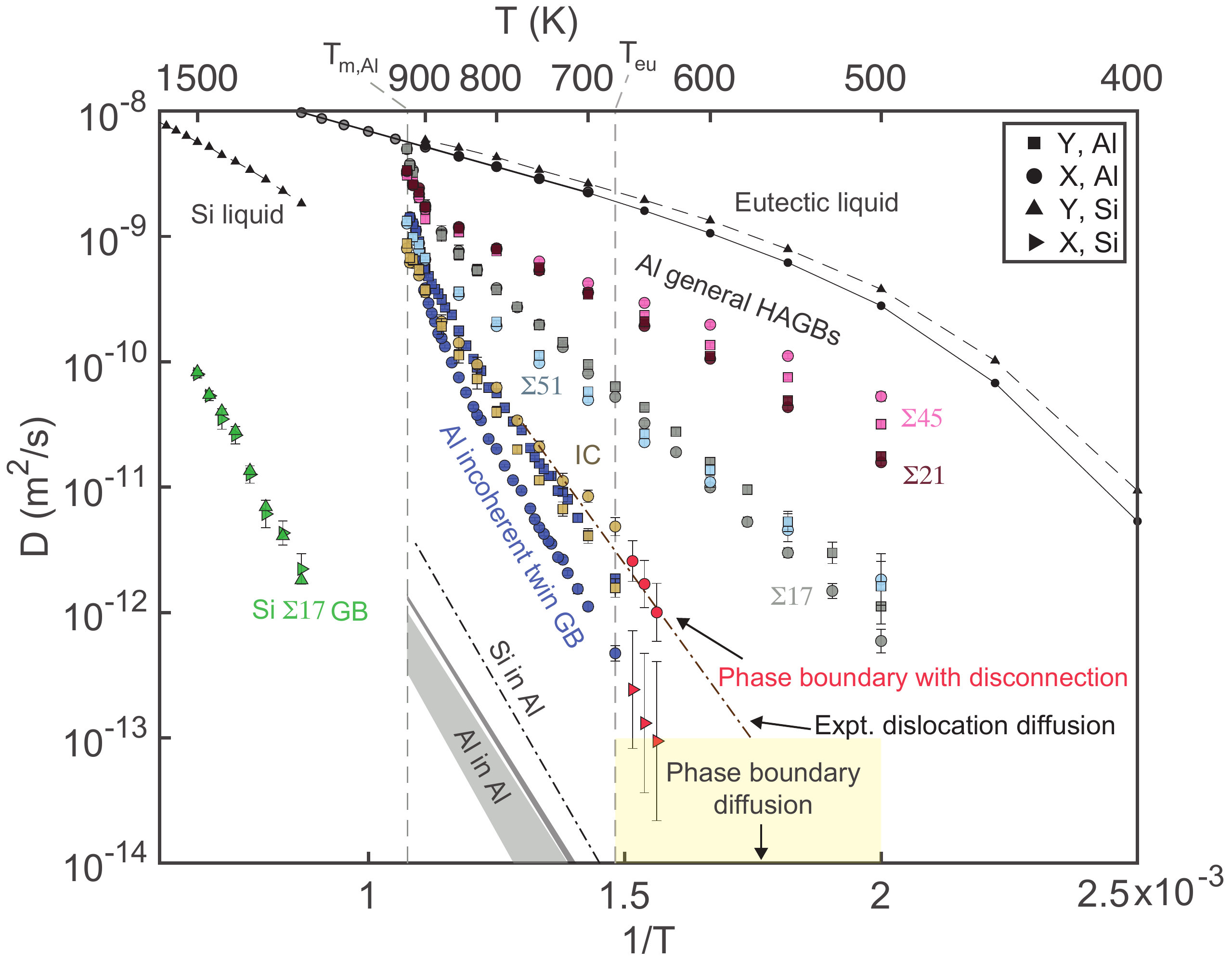}
\caption{Arrhenius diagram of diffusion in Al, Si, and Al-Si systems. The yellow
box indicates the estimated range of diffusivity in planar or faceted
Al-Si interphase boundaries without extrinsic defects. The Al-Si interface
containing disconnections (red symbols) exhibits Al diffusion rates
lower than those for general high-angle GBs (HAGBs) and similar to
those for the incommensurate (IC) and incoherent twin GBs. Many GBs
exhibit anisotropic diffusion in the $X$ and $Y$ directions. Experimental
data is shown for lattice diffusion (Al in Al, Si in Al) and Si diffusion
along Al dislocations (dash-dotted line).}
\label{fig:fig6}
\end{figure}

Diffusion coefficients in the Al-Si interphase boundaries and several
reference systems are shown on the Arrhenius diagram (log diffusivity
versus inverse temperature) in Fig.~\ref{fig:fig6}. The reference
systems include pure Al and pure Si GBs, pure Al and pure Si liquids,
the eutectic liquid, and a Si-enriched Al GB. We also include experimental
data for indirect measurements of Si dislocation diffusion in Al \citep{legros_expt}.
Experimental data and DFT calculations are also shown for lattice
diffusion in pure Al and Si diffusion in Al-Si alloys. Table \ref{tab:T2}
summarizes the diffusion activation energies and prefactors for each
system. For the Al-Si interface with disconnections, the phase compositions
vary with temperature along the solvus lines on the phase diagram.
Thus, the temperature dependence of the Al and Si diffusivities is
not expected to follow the Arrhenius law. Given also that the calculations
include only three temperatures in a narrow temperature interval,
we refrained from predicting the Arrhenius parameters in this case.

The main conclusion from Fig.~\ref{fig:fig6} is that the Al-Si interphase
boundaries suppress Al diffusion relative to Al GBs and accelerate
Si diffusion relative to Si GBs. The comparison with GBs requires
an explanation because GB diffusivities span several orders of magnitude,
depending on the GB type. Out of the six Al GBs studied in this work,
the $\Sigma3\;(211)$ incoherent twin boundary and the $(100)||(110)$
incommensurate GB have the lowest diffusion coefficients, the highest
activation energies, and do not exhibit any significant pre-melting.
They represent the lower envelope of diffusion coefficients in high-angle
GBs in Al and can be used for comparison with interphase boundaries.
The incommensurate GB has the same macroscopic crystallography as
the $\left\{ 110\right\} _{\mathrm{Al}}||\left\{ 001\right\} _{\mathrm{Si}}$
interphase boundary. The interphase boundary produced by simulated
epitaxy is atomically flat and has such a low diffusivity that it
could not be resolved by MD simulations. It certainly lies below the
diffusivity of even the slowest-diffusion GBs in Al. The direct bonding
method produces a metastable structure of this interface that contains
disconnections, which increase the rates of both Al and Si diffusion.
This disconnection-accelerated interface diffusivity is comparable
with the Al diffusivity in the incommensurate and incoherent twin
GBs. In other words, it reaches the lower bound of Al GB diffusivities. 

\begin{figure}[H]
\centering{}\includegraphics[width=0.7\textwidth]{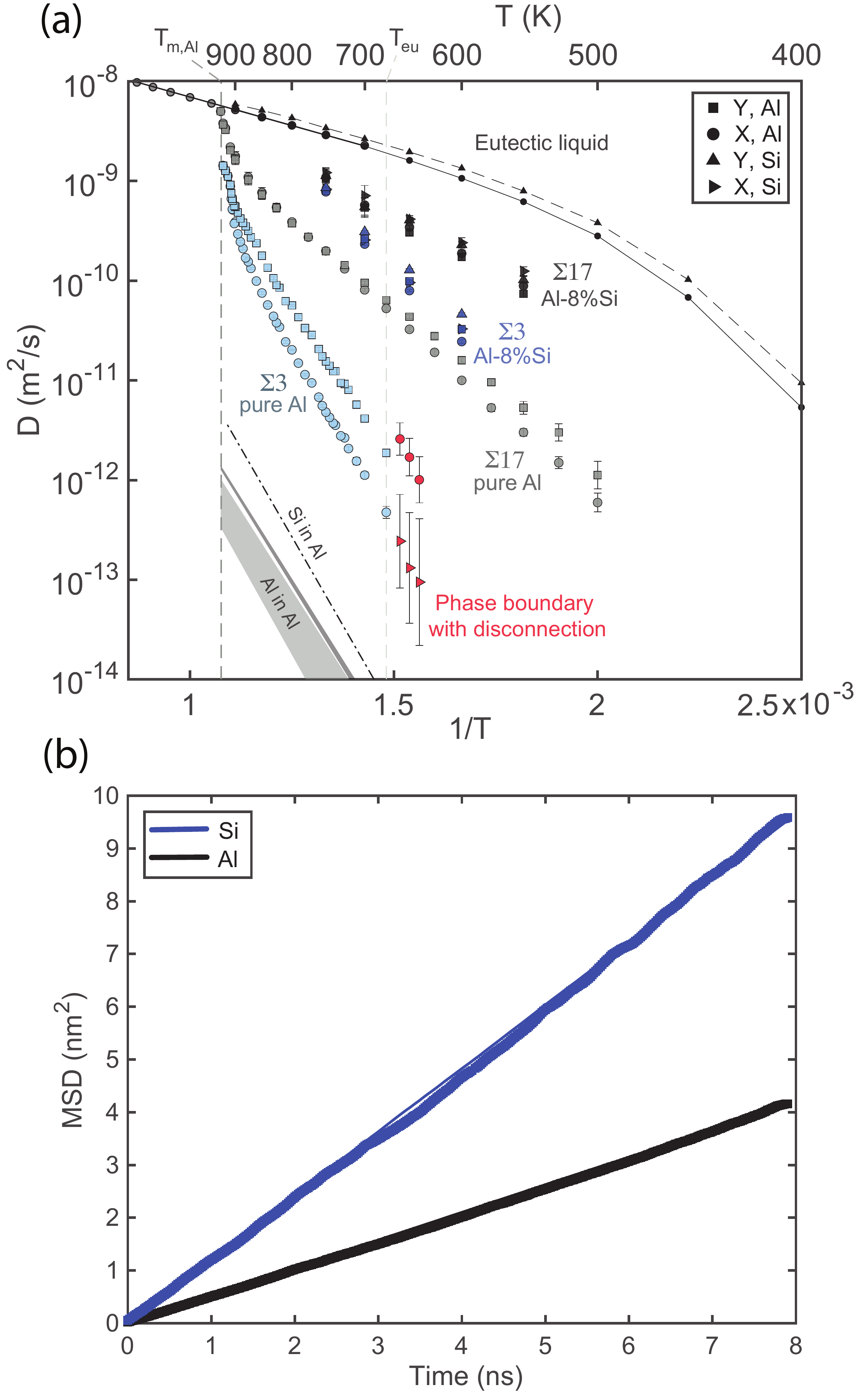}
\caption{(a) Arrhenius diagram of Al and Si diffusion in the Si-enriched $\Sigma3$
and $\Sigma17$ GBs in Al-8at.\%Si alloy. Selected data from Fig.~\ref{fig:fig6}
is included for comparison. (b) Example of MSD versus time curves
along the tilt axis of the $\Sigma3$ incoherent twin GB at 650 K.
The lines represent linear fits. The plots demonstrate that Si atoms
diffuse in GBs faster than Al atoms.}
\label{fig:fig7}
\end{figure}

\begin{figure}[H]
\centering{}\includegraphics[width=0.6\textwidth]{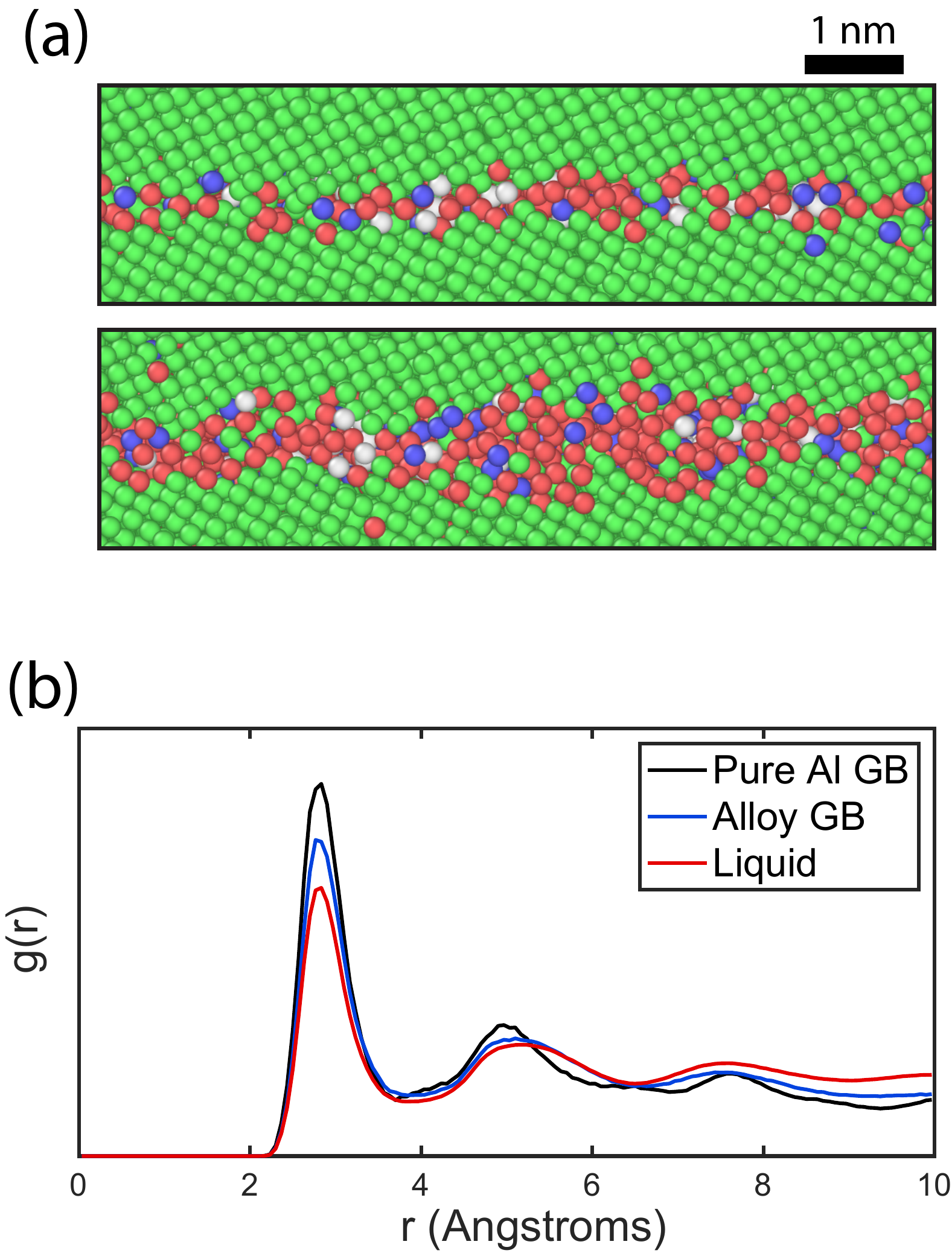}
\caption{GB structures in Al-Si alloys at the temperature of 750 K. (a) $\Sigma17$
GB in pure Al (top panel) and Al-8at.\%Si alloy (bottom panel). The
atoms are colored by local coordination: green = FCC, blue = DC, and
all other colors represent atoms in locally disordered environments.
(b) Radial distribution function of Al atoms in the core of the pure
Al GB, the alloy GB, and in supercooled eutectic liquid. The plot
demonstrates that the alloy GB is in intermediate state of order between
the Al GB and bulk liquid.}
\label{fig:fig8}
\end{figure}

The results shown in Fig.~\ref{fig:fig6} indicate that Si diffusivity
tends to increase in the order: Si crystal $\rightarrow$ Si GBs $\rightarrow$
Si in Al lattice $\rightarrow$ Si in planar Al-Si interface $\rightarrow$
Si in Al-Si interface with disconnections $\rightarrow$ Si in Al
GBs $\rightarrow$ Si in Si-Al liquid. In particular, the extrapolated
rate of Si diffusion in the Si $\Sigma17\;\left\langle 100\right\rangle $
GB falls well below Si diffusivity in the Al-Si interface with disconnections
and likely below the Si diffusivity in the interface without disconnections.
To evaluate the Si diffusivity in Al GBs, we used a Si-enriched Al
GB obtained by interface-induced recrystallization when the new GB
remained immobile after splitting from the Al-Si interface. The Al
and Si diffusion coefficients in this GB fall in the upper range of
GB diffusion coefficients in high-angle Al GBs. A more systematic
test was performed by studying diffusion in the $\Sigma17$ and $\Sigma3$
GBs in Al-8at.\%Si alloy (Fig.~\ref{fig:fig7}a). Si atoms were observed
to strongly segregate to these GBs and form a 1 nm or thicker disordered
(liquid-like) layer with eutectic composition (Fig.~\ref{fig:fig8}a,
bottom panel). The liquid-like nature of the segregated layer was
confirmed by the shift of the radial distribution function toward
the one for the eutectic liquid (Fig.~\ref{fig:fig8}a). Both Al
and Si were found to diffuse significantly faster in the segregated
GBs than Al self-diffusion in pure Al GBs with the same crystallography
(Fig.~\ref{fig:fig7}a). Si is observed to diffuse in the segregated
GBs slightly faster than Al (\ref{fig:fig7}b), mirroring a similar
trend in crystalline Al and eutectic liquid. In all these cases, the
Si diffusivity far exceeds its diffusivity in Al-Si interfaces.

Finally, we note that diffusion coefficients in the eutectic liquid
show a downward deviation from the Arrhenius law (Fig.~\ref{fig:fig6})
reminiscent of that in fragile glass-forming liquids \citep{AlSm1}.
As in the studies of model fragile metallic glasses \citep{derlet2021viscosity},
percolating networks of icosahedrally coordinated atoms were apparent
in the structure of the deeply supercooled Al-Si liquid. Crystallization
of the eutectic liquid was observed at sufficiently low temperatures
(around 0.6 $T_{\mathrm{eu}}$).

\section{Discussion\label{sec:Discussion}}

Over the past two decades, new atomistic methods have been developed
to accurately calculate GB diffusion coefficients and better understand
atomic-level GB diffusion mechanisms \citep{Sorensen00,Suzuki03a,Suzuki04a,Suzuki05a,gbmdhpnas,Frolov2013a,Cu-Ag-segregation-diffusion,Koju:2020ab,mishin2015atomistic,collective_Fe,Chesser:2022aa}.
To our knowledge, this work is the first application of such methods
to diffusion along metal-nonmetal interphase boundaries.

The main discovery of this work is that the diffusion mobilities of
Al and Si atoms in stable Al-Si interphase boundaries are much lower
than their mobilities in Al GBs (Fig.~\ref{fig:fig6}). They are
also lower than the mobilities in interphase boundary disconnections
existing in metastable interface structures. In fact, the diffusivity
in stable planar interphase boundaries is so low that we could not
quantify it by MD simulations. We could only estimate an upper bound
of possible interface diffusion coefficients (Fig.~\ref{fig:fig6}). 

We hypothesize that the sluggish interphase boundary diffusion is
a common feature of interfaces between highly dissimilar materials
with significantly different chemical bond strengths. In our case,
Si is a covalently bonded material with a significantly higher melting
temperature and greater mechanical strength than metallic Al. Al-Si
interfaces can only exist below the eutectic temperature $T_{\mathrm{eu}}=850$
K \citep{EXPT_phase_dia}, which corresponds to the homologous temperature
of 0.91 for Al and 0.50 for Si. In other words, the eutectic temperature
is very hot for Al and very cold for Si. Accordingly, at temperatures
close to $T_{\mathrm{eu}}$, the mobility of Al atoms is very high
in both Al GBs and FCC lattice, while the Si mobility is drastically
lower in both Si GBs and DC lattice. Although the interatomic potential
used in this work underestimates the experimental eutectic temperature,
it correctly reproduces this basic feature of the Al-Si system, including
the large mobility gap between Al and Si.

The Si atoms residing on the Si side of the Al-Si interfaces are virtually
immobile on the diffusion timescale of Al atoms. The strongly ordered
Si positions impose a periodic potential that restrains the motion
of both Al and Si atoms along the interface, suppressing the interface
diffusivity. By contrast, both Al and Si atoms diffusing along Al
GBs are free from the mentioned constraint, leading to higher mobility
relative to the interphase boundaries (Figs.~\ref{fig:fig6} and
\ref{fig:fig7}). For the same reason, at temperatures on the order
of $T_{\mathrm{eu}}$, Si diffusivity in FCC Al is much higher than
in the DC Si lattice, where the atoms are subject to the strong ordering
constraint. This explanation of the sluggishness of the Al-Si interface
diffusion is purely qualitative and requires a more rigorous quantitative
formulation in the future.

Even when the interface diffusivity was too low to calculate the diffusion
coefficients, diffusion mechanisms could still be observed. We find
that the interface atoms diffuse predominantly by collective rearrangements,
often in the form of strings or rings of correlated displacements.
Similar mechanisms were previously observed in the simulations of
GB self-diffusion in elemental metals \citep{Sorensen00,Suzuki03a,Suzuki04a,Suzuki05a,mishin2015atomistic,collective_Fe,Chesser:2022aa}.
In this work, we observed the operation of the collective mechanisms
during the co-diffusion of two atomic species, Al and Si. Both species
participate in the collective rearrangements, but to a different extent
due to the difference in their mobilities. Si atoms participate in
collective events to a lesser extent than Al atoms. It also appears
that they tend to disrupt the chains of Al displacements relative
to those in pure Al environments.

One of the challenges in modeling interphase boundaries is constructing
a two-phase system in full thermodynamic equilibrium and finding the
orientation relationships corresponding to the most stable interface
structures. This task requires close integration of MD and MC methodologies
and constituted a significant part of the present work. One of the
unexpected findings was the phenomenon of interface-induced recrystallization.
An interface brought to a non-equilibrium state by thermal processing
or deformation can transform to a more stable state by changing the
orientation relationship between the phases and injecting a new grain
into the metallic phase. This process and its possible impact on the
mechanical properties of metal-ceramic composites is worth further
exploration by experiment and modeling in the future.

\section{Conclusions}

We have studied the atomic structure, structural stability, and diffusion
processes in Al-Si interphase boundaries by combining MD and MC simulations
with an interatomic potential \citep{saidi_aeam}. Two different interface
preparation methods were applied: direct bonding and simulated epitaxy.
Diffusion coefficients in the interphase boundaries were calculated
and compared with those in Al GBs and bulk liquid. Interface diffusion
mechanisms were investigated by statistical methods developed previously
for GBs and supercooled glass-forming liquids. The following conclusions
can be drawn from this work.

(1) The most stable interphase boundaries were found to be either
planar or nano-faceted and tended to have either cubic or twin-related
orientation relationships. Many of the orientation relationships found
in this work were previously observed in experiments. The stable interfaces
did not pre-melt and could be overheated above the eutectic temperature. 

(2) Diverse metastable interface structures were observed, depending
on the fabrication method and lattice misfit. Examples include interfaces
with disconnections and interfaces shooting out threading dislocations
into the metallic phase. Disconnections were found to accelerate interface
diffusion, while the facet edges did not. The impact of other interface
defects, such as absorbed lattice dislocations, on interface diffusion
and other properties of Al-Si composites is yet to be studied.

(3) The simulations have revealed an interesting phenomenon of interface-induced
recrystallization of the metallic phase. The new metallic grains can
be nanometer-scale thin or can grow deep into the metallic phase.
The moving GBs bounding the growing grain carry along some of the
Si atoms and can be slowed down by the solute drag effect. We envision
that interface-induced recrystallization may occur in metal-matrix
composites during high-temperature creep deformation. The effect resembles
the diffusion-induced recrystallization previously observed in diverse
systems \citep{chae1996diffusion,schmitz2010hidden,levi2022diffusion}. 

(4) Diffusion of Al and Si atoms in stable Al-Si interfaces is surprisingly
slow and could not be calculated on the MD timescale accessible to
us. We estimate that the interface diffusivities are below $D=10^{-13}$
m$^{2}$/s even at the eutectic temperature. The interphase boundary
diffusion is much slower than diffusion of both Al and Si in Al GBs.
On the other hand, disconnections existing in metastable interface
structures greatly accelerate diffusion to a level where it can be
readily calculated and was found to be close to the diffusivity in
some of the Al GBs. We suggest that this observation could be relevant
to the loss of creep resistance of composites at high temperatures.
If the initial interface sliding creates a set of interface dislocations
and/or disconnections, the accelerated interface diffusion will promote
further sliding, triggering a self-amplified process that eventually
causes the loss of sliding resistance. 

(5) The sluggish interphase boundary diffusion could be a common feature
of many metal-ceramic systems with significantly different bond strengths,
melting temperatures, and thus diffusivities in the two phases. The
highly ordered atomic structure of the ceramic phase can suppress
the diffusion mobility in the interface region relative to the less
constrained atomic environments in metallic GBs. This can qualitatively
explain the faster GB diffusion than diffusion along the interphase
boundaries. In the Al-Si system, the additional effect is the disordering
of Al GBs caused by Si segregation, resulting in further acceleration
of GB diffusion.

(6) Similar to GB diffusion, diffusion in interphase boundaries is
dominated by correlated atomic rearrangements involving strings and
rings of simultaneously moving atoms. Both Al and Si atoms participate
in the collective mechanisms, although to a different extents. Collective
mechanisms of interface co-diffusion of two or more species with different
mobilities is an interesting topic for future research.

(7) The interatomic potential \citep{saidi_aeam} used in this work
reproduces the Al-Si phase diagram qualitatively correctly but underestimates
the eutectic temperature and overestimates the solubility limits in
the solid Al and Si. A more accurate interatomic potential for this
system should be developed to ensure quantitatively accurate simulation
results.

\bigskip{}

\noindent \textbf{Acknowledgements}

\noindent This research was supported by the U.S.~Department of Energy,
Office of Basic Energy Sciences, Division of Materials Sciences and
Engineering, under Award \# DE-SC0023102.

\bibliographystyle{unsrt}
\bibliography{citations}

\newpage{}
\begin{table}[H]
\centering %
\begin{tabular*}{1\textwidth}{@{\extracolsep{\fill}}@{\extracolsep{\fill}}lllll}
\toprule 
Si (hkl) & Processing & Al-Si OR & Disorientation & Reconstructions\tabularnewline
\midrule 
(1 1 1) & Epitaxy & Cube & $[-0.22,0.77,0.59]$ 0.1$^{\circ}$ & \tabularnewline
 & Direct bonding & Cube & 0$^{\circ}$ & \tabularnewline
(1 0 0) & Epitaxy & Twin-related & $[-0.68,-0.68,0.28]$ 62.8$^{\circ}$ & Threading screw\tabularnewline
 &  &  &  & dislocations\tabularnewline
 & Direct bonding & Twin-related & $[-0.68,0.28,0.68]$ 62.8$^{\circ}$ & Disconnections\tabularnewline
 & Direct bonding & Cube & 0$^{\circ}$ & Nano-facets\tabularnewline
(1 1 0) & Epitaxy & Twin-related & $[1,0,-1]$ 59.8$^{\circ}$ & \tabularnewline
 & Direct bonding & Cube & 0$^{\circ}$ & \tabularnewline
(1 1 2) & Epitaxy & Cube-related & $[-0.62,0.71,-0.26]$ 6.2$^{\circ}$ & \tabularnewline
 & Recrystallization & Cube-related & $[0.63,0.69,-0.37]$ 4.7 $^{\circ}$ & \tabularnewline
 & Direct bonding & Cube & 0$^{\circ}$ & Stacking faults extended\tabularnewline
 &  &  &  & from interface\tabularnewline
(3 1 0) & Epitaxy & Cube-related & $[0.7,0.61,-0.36]$ 11.3$^{\circ}$ & \tabularnewline
 & Recrystallization & Cube-related & $[310]$ 0.2 $^{\circ}$ & Nano-facets\tabularnewline
(2 1 0) & Epitaxy & Cube-related & $[-0.72,-0.46,-0.52]$ 11.3$^{\circ}$ & \tabularnewline
 & Recrystallization & Cube-related & $[0.34,0.8,-0.49]$ 1.9 $^{\circ}$ & Nano-facets\tabularnewline
(8 5 1) & Epitaxy & Cube-related & $[0.96,-0.18,0.21]$ 8.5$^{\circ}$ & \tabularnewline
 & Direct bonding & Cube & 0$^{\circ}$ & \tabularnewline
(1 2 3) & Epitaxy & Cube-related & $[0.63,0.12,0.77]$ 8.9$^{\circ}$ & \tabularnewline
(1 1 5) & Epitaxy & Cube-related & $[1,1,0]$ 11.1$^{\circ}$ & \tabularnewline
(3 3 2) & Epitaxy & Cube-related & $[-0.3,-0.27,-0.91]$ 2.8$^{\circ}$ & \tabularnewline
\bottomrule
\end{tabular*}\caption{Stable Al-Si interfaces studied in this work. The table shows Miller
indices $(hkl)$ of the Si phase, disorientation (indices and angle)
of the Al phase, and the type of crystallographic orientation relationship
(OR) between the two. \label{tab:T1}}
\end{table}

\newpage{}
\begin{table}[H]
\begin{tabular}{lccc}
\toprule 
System & Direction & $E$ (eV) & $D_{0}$ (m$^{2}$/s)\tabularnewline
\midrule
Si in Al dislocation$^{*}$ \citep{legros_expt} &  & 1.12 & 7$\times10^{-4}$\tabularnewline
\midrule 
Al grain boundaries: &  &  & \tabularnewline
Al(110) $\parallel$ Al(100) & 110 $\parallel$ 110 & 0.87 $\pm$ 0.06 & (1.4$_{-2.1}^{+2.9}$ )$\times10^{-5}$\tabularnewline
 & 100 $\parallel$ 110 & 1.10 $\pm$ 0.07 & (3.1$_{-5.9}^{+8.0}$)$\times10^{-4}$\tabularnewline
$\Sigma3$ $\left\langle 110\right\rangle $ tilt 70.5$^{\circ}$ & $\parallel$ tilt axis & 1.31 $\pm$ 0.03 & (3.89$_{-1.2}^{+2.1}$)$\times10^{-3}$\tabularnewline
 & $\perp$ tilt axis & 1.45 $\pm$ 0.03 & (1.17$_{-2.7}^{+4.8}$)$\times10^{-2}$\tabularnewline
$\Sigma17$ $\left\langle 100\right\rangle $ tilt 61.9$^{\circ}$ & $\parallel$ tilt axis & 0.72 $\pm$ 0.01 & (5.53$_{-0.6}^{+1.1}$)$\times10^{-6}$\tabularnewline
 & $\perp$ tilt axis & 0.81 $\pm$ 0.01 & (2.09$_{-0.17}^{+0.33}$)$\times10^{-5}$\tabularnewline
$\Sigma51$ $\left\langle 551\right\rangle $ tilt 180$^{\circ}$ & average & 0.55 $\pm$ 0.02 & (4.5$_{-1.2}^{+2.2}$)$\times10^{-7}$\tabularnewline
$\Sigma21$ $\left\langle 531\right\rangle $ asymmetric tilt 80.4$^{\circ}$ & average & 0.44 $\pm$ 0.01 & (5.4$_{-0.8}^{+1.6}$)$\times10^{-7}$\tabularnewline
$\Sigma45$ $\left\langle 851\right\rangle $ tilt 180$^{\circ}$ & average & 0.34 $\pm$ 0.01 & (9.7$_{-0.9}^{+1.8}$)$\times10^{-8}$\tabularnewline
\midrule 
Si grain boundaries: &  &  & \tabularnewline
$\Sigma17$ $\left\langle 100\right\rangle $ tilt 61.9$^{\circ}$ & average & 1.59 $\pm$ 0.05 & (1.9$_{-1.0}^{+1.7}$)$\times10^{-5}$\tabularnewline
\midrule 
Solids: &  &  & \tabularnewline
Al in Al$^{*}$ \citep{EXPT_Al_bulk_diffusion_data} &  & 1.32 & 1.8 $\times10^{-5}$\tabularnewline
Al in Al$^{^{**}}$ \citep{DFT_Al_bulk_diffusion_data} &  & 1.25$-$1.35 & (0.542$-$2.42) $\times10^{-5}$\tabularnewline
Si in Al$^{*}$ \citep{Si_in_Al,Du:2003aa} &  & 1.41 & 2 $\times10^{-4}$\tabularnewline
Si in Si, vacancy mechanism$^{*}$ \citep{Sudkamp:2016aa} &  & 4.65 & 1.1 $\times10^{-2}$\tabularnewline
Si in Si, interstitialcy mechanism$^{*}$ \citep{Sudkamp:2016aa} &  & 4.83 & 5.5 $\times10^{-2}$\tabularnewline
Si in amorphous Si$^{*}$ \citep{aSi_diffusion} &  & 4.4 $\pm$ 0.3 & 1.5 $\times10^{5}$\tabularnewline
\midrule 
Liquids: &  &  & \tabularnewline
Al in liquid Al &  & 0.227 $\pm$ 0.001 & (9.69$_{-0.12}^{+0.24}$)$\times10^{-8}$\tabularnewline
Al in eutectic liquid &  & 0.231 $\pm$ 0.001 & (1.03$_{-0.01}^{+0.03}$)$\times10^{-7}$\tabularnewline
Si in eutectic liquid &  & 0.218 $\pm$ 0.007 & (1.0$_{-0.09}^{+0.17}$)$\times10^{-7}$\tabularnewline
Si in Si liquid &  & 0.429 $\pm$ 0.007 & (1.60$_{-0.09}^{+0.16}$)$\times10^{-7}$\tabularnewline
\bottomrule
\end{tabular}\caption{Activation energies ($E$) and prefactors ($D_{0}$) for diffusion
in Al-Si systems calculated in this work. Shown for comparison is
literature data for $^{*}$experiment and $^{**}$DFT calculations.}
\label{tab:T2}
\end{table}

\newpage{}

\global\long\def\figurename{Supplementary Figure}%
\global\long\def\tablename{Supplementary Table}%
\setcounter{figure}{0} \setcounter{table}{0}

\section*{Supplementary Information}

\begin{table}[H]
\centering %
\begin{tabular}{lcccc}
\toprule 
GB description & $X$ & $Y$ & $Z$ & $\gamma$ (mJ/m$^{2}$)\tabularnewline
\midrule 
\multirow{2}{*}{$\Sigma3$ $\left\langle 110\right\rangle $ tilt 70.5$^{\circ}$} & $[-1,1,1]$ & $[0,1,-1]$ & $[2,1,1]$ & \multirow{2}{*}{418}\tabularnewline
 & $[1,-1,-1]$ & $[0,-1,1]$ & $[2,1,1]$ & \tabularnewline
 &  &  &  & \tabularnewline
\multirow{2}{*}{Al(110) $\parallel$ Al(100)} & $[0,0,1]$ & $[1,-1,0]$ & $[1,1,0]$ & \multirow{2}{*}{389}\tabularnewline
 & $[1,1,0]$ & $[-1,1,0]$ & $[0,0,1]$ & \tabularnewline
 &  &  &  & \tabularnewline
\multirow{2}{*}{$\Sigma51$ $\left\langle 551\right\rangle $ tilt 180$^{\circ}$} & $[5,-5,1]$ & $[8,9,5]$ & $[-2,-1,5]$ & \multirow{2}{*}{498}\tabularnewline
 & $[5,5,-1]$ & $[-8,9,5]$ & $[2,-1,5]$ & \tabularnewline
 &  &  &  & \tabularnewline
\multirow{2}{*}{$\Sigma17$ $\left\langle 100\right\rangle $ tilt 61.9$^{\circ}$} & $[5,-3,0]$ & $[0,0,-1]$ & $[3,5,0]$ & \multirow{2}{*}{488}\tabularnewline
 & $[-5,3,0]$ & $[0,0,1]$ & $[3,5,0]$ & \tabularnewline
 &  &  &  & \tabularnewline
\multirow{2}{*}{$\Sigma21$ $\left\langle 531\right\rangle $ asymmetric tilt 80.4$^{\circ}$} & $[10,1,-5]$ & $[1,5,3]$ & $[4,-5,7]$ & \multirow{2}{*}{569}\tabularnewline
 & $[5,-3,-1]$ & $[2,3,1]$ & $[0,-1,3]$ & \tabularnewline
 &  &  &  & \tabularnewline
\multirow{2}{*}{$\Sigma45$ $\left\langle 851\right\rangle $ tilt 180$^{\circ}$} & $[11,2,1]$ & $[-1,8,-5]$ & $[-1,3,5]$ & \multirow{2}{*}{593}\tabularnewline
 & $[11,-2,1]$ & $[1,8,5]$ & $[-1,-3,5]$ & \tabularnewline
\bottomrule
\end{tabular}\caption{Crystallography and 0 K energies of Al GBs studied in this work. The
Cartesian $X$, $Y$ and $Z$ axes are aligned with edges of the rectangular
simulation box. The GB plane is normal to the $Z$-direction. For
each GB, the two lines indicate crystallographic directions parallel
to the Cartesian axes in the upper and lower grains. The 0 K GB energy
($\gamma$) was computed with the interatomic potential from Ref.~\citep{Almendelev}.
The data is sorted based on the magnitude of the diffusion coefficient
at 700 K from lowest to highest. \label{tab:S1}}
\end{table}

\begin{figure}[H]
\centering\leavevmode \includegraphics[width=0.7\textwidth]{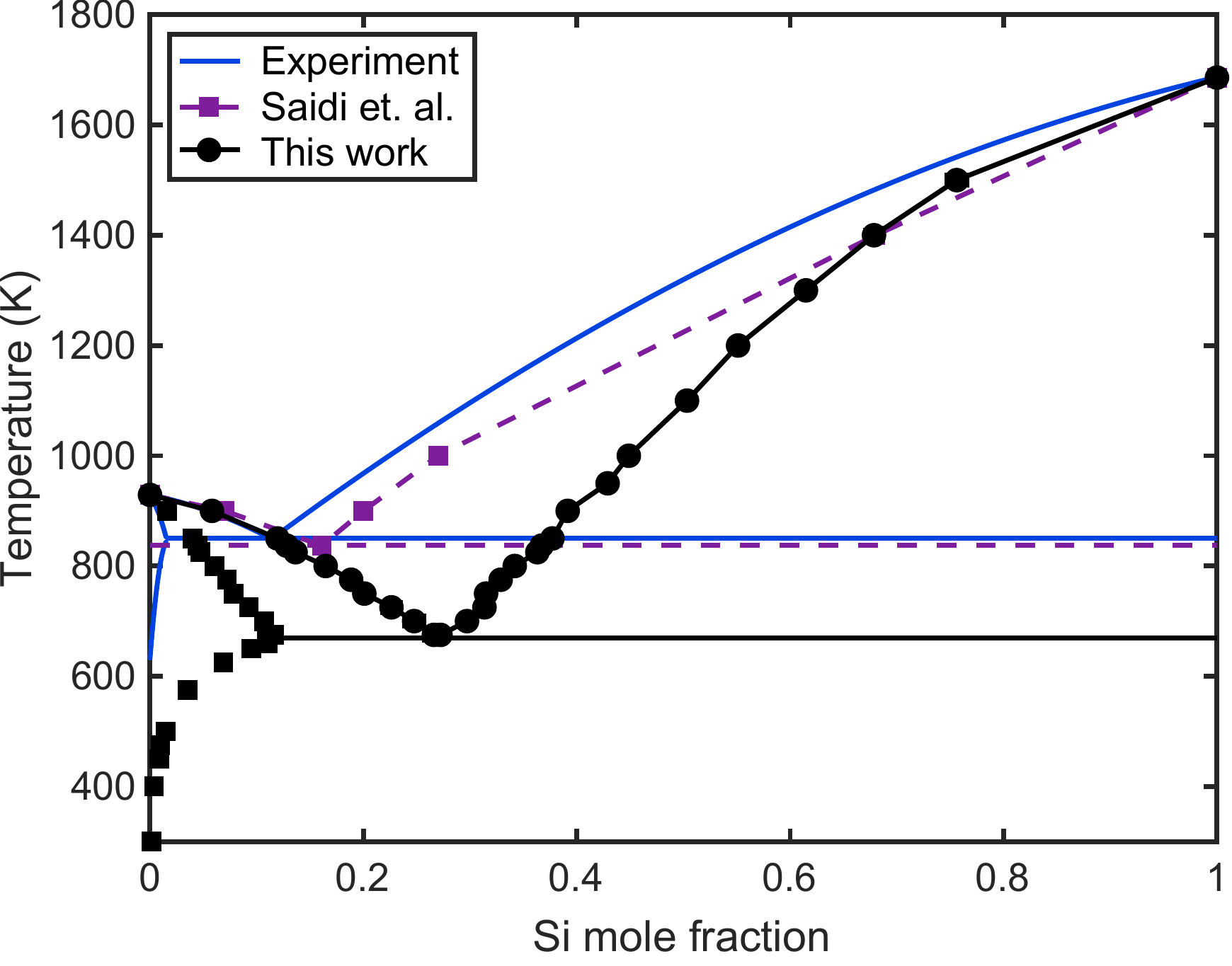}
\caption{Al-Si phase diagram computed in this work in comparison with the experimental
diagram \citep{EXPT_phase_dia} and previous calculations by Saidi
et al.~\citep{saidi_aeam}.}
\label{fig:figS1}
\end{figure}

\begin{figure}[H]
\centering\leavevmode \includegraphics[width=0.8\textwidth]{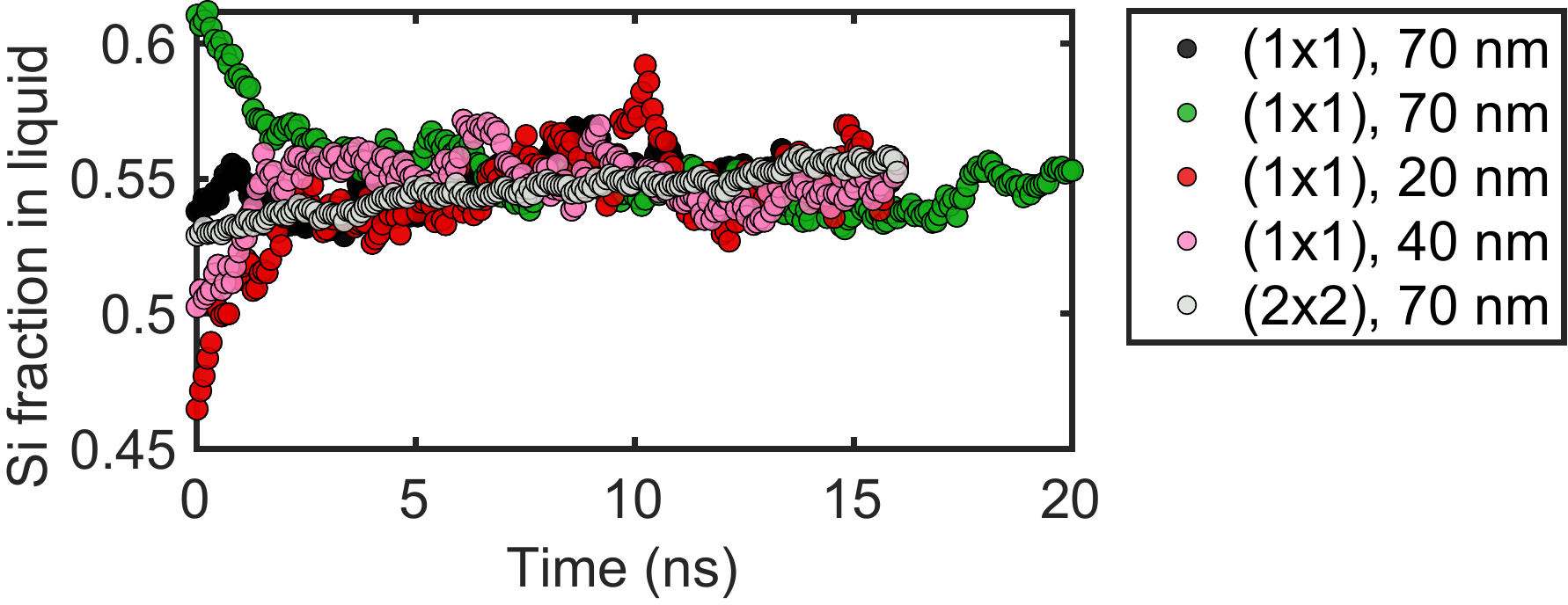}
\caption{Examples of a convergence checks for MC-MD calculations of the Al-Si
phase diagram. Data is shown for an example point on the liquidus
line on the Si-rich side of the diagram for multiple initial Si concentrations
in the liquid and multiple solid-liquid interface cross-section sizes.
In the legend, (1x1) refers to the integer repeats of a 1.2 nm x 1.2
nm interface cross section, while the second number is the thickness
for the liquid and solid layers.}
\label{fig:figS2}
\end{figure}

\begin{figure}[H]
\centering\leavevmode \includegraphics[width=1\textwidth]{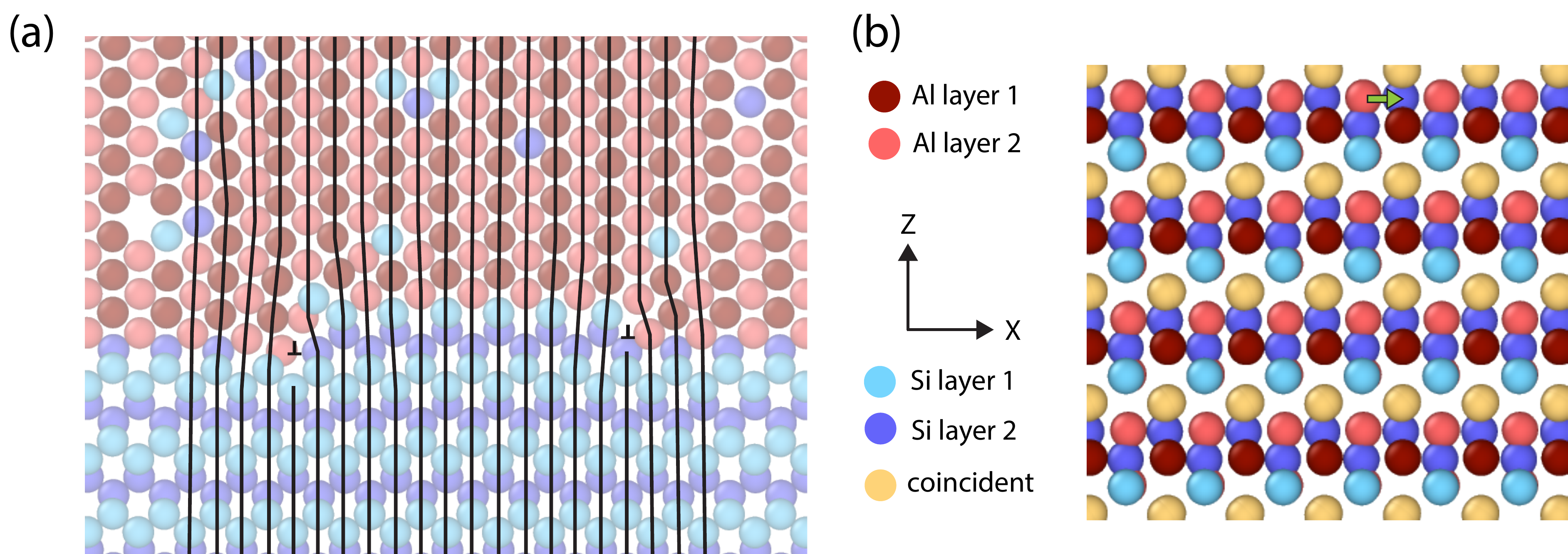}
\caption{Characterization of phase boundary disconnections. (a) The $\left\{ 110\right\} _{\mathrm{Al}}||\left\{ 001\right\} _{\mathrm{Si}}$
interface structure with the same color coding as in Fig.~\ref{fig:fig3}a
of the main text. Vertical atomic rows in coherent regions are outlined
to reveal the dislocation content of the interface localized near
the terminated atomic planes. (b) Coherent dichromatic pattern of
the same interface obtained by appropriate straining of the Al phase.
The green arrow indicates the disconnection Burgers vector, whose
indices are $1/2\left\langle 100\right\rangle $ relative to the Al
lattice and $1/2\left\langle 110\right\rangle $ relative to the Si
lattice.}
\label{fig:figS5}
\end{figure}

\section*{Further study of interface-induced recrystallization }

\begin{figure}[H]
\centering{}\includegraphics[width=1\textwidth]{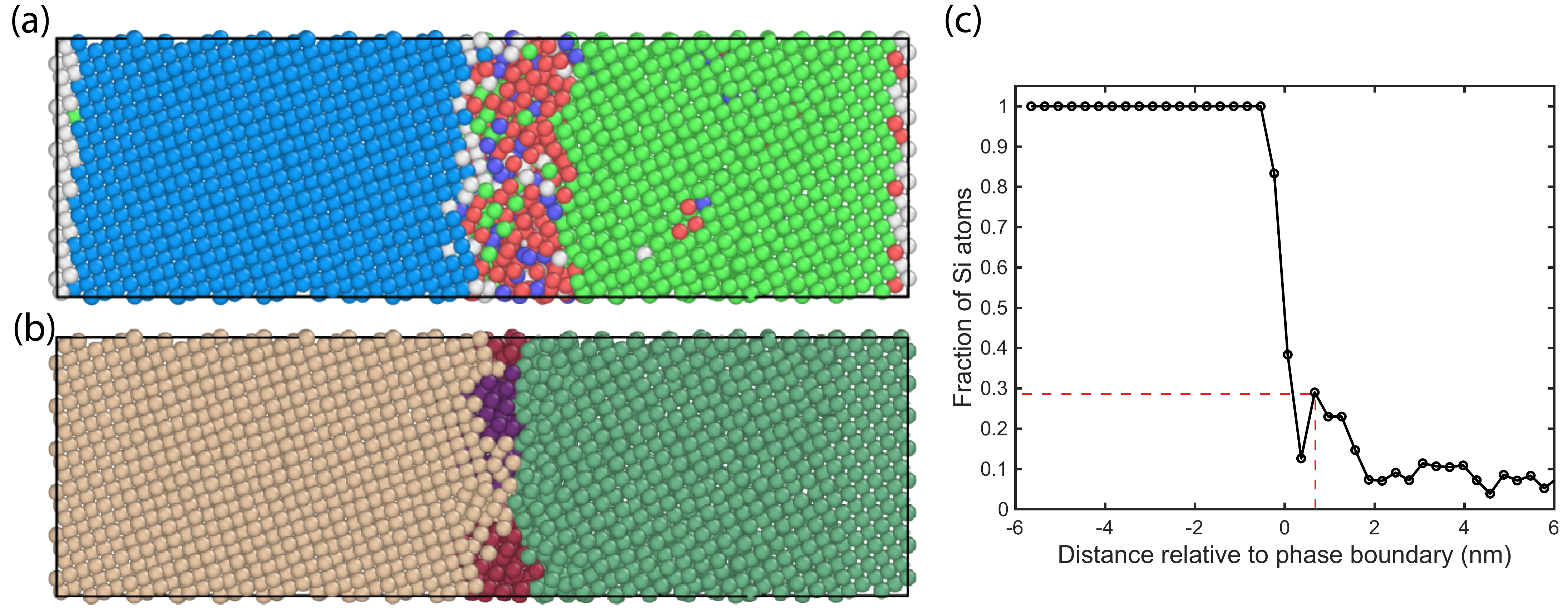}
\caption{Example of interface-induced recrystallization without GB migration.
The initial interphase boundary has the same crystallography as a
$\Sigma5$ (210) $\left\langle 100\right\rangle $ tilt GB. (a) After
an MC/MD anneal, the interface develops a Si-enriched, partially disordered
layer on the Al side. (b) Spontaneous formation and dissolution of
subcritical nuclei of new grains is revealed by grain segmentation
analysis (distinct grains have different colors). (c) The Si concentration
peak with a near eutectic composition (dashed lines) can be interpreted
as a new GB with Si segregation. This splitting of the interface structure
represents an embryonic form of interface-induced recrystallization.}
\label{fig:figS3}
\end{figure}

Al recrystallization induced by the Al-Si interphase boundary starts
with a transient state in which a Si-enriched layer forms on the Al
side of the interface. This layer can be interpreted as a nucleus
of a new Al grain separated from the parent grain by a GB containing
some amount of segregated Si. How far this GB migrates into the parent
grain depends on whether a sufficiently large thermodynamic force
exists for GB migration. We hypothesize that, in our simulations,
the driving force is an elastic strain energy difference across the
GB which arises due to the lattice misfit at the Al-Si interface.
A large strain energy difference can drive the growth of the new grain
deep into the old. If, by crystal symmetry, there is no elastic energy
difference across the GB, as in the case of symmetric tilt GBs, or
if the misfit strain energy density is too small, then the driving
force will be insufficient to cause significant GB migration. The
recrystallization will then results in a narrow splitting of the initial
interface into one with a more stable crystallographic orientation
and a thin layer representing a new Al grain parallel to the interface.
The latter scenario is illustrated in Fig.~\ref{fig:figS3}, where
the recrystallization does not proceed beyond creating a Si-enriched
GB on the Al side.

\begin{figure}[H]
\centering\leavevmode \includegraphics[width=1\textwidth]{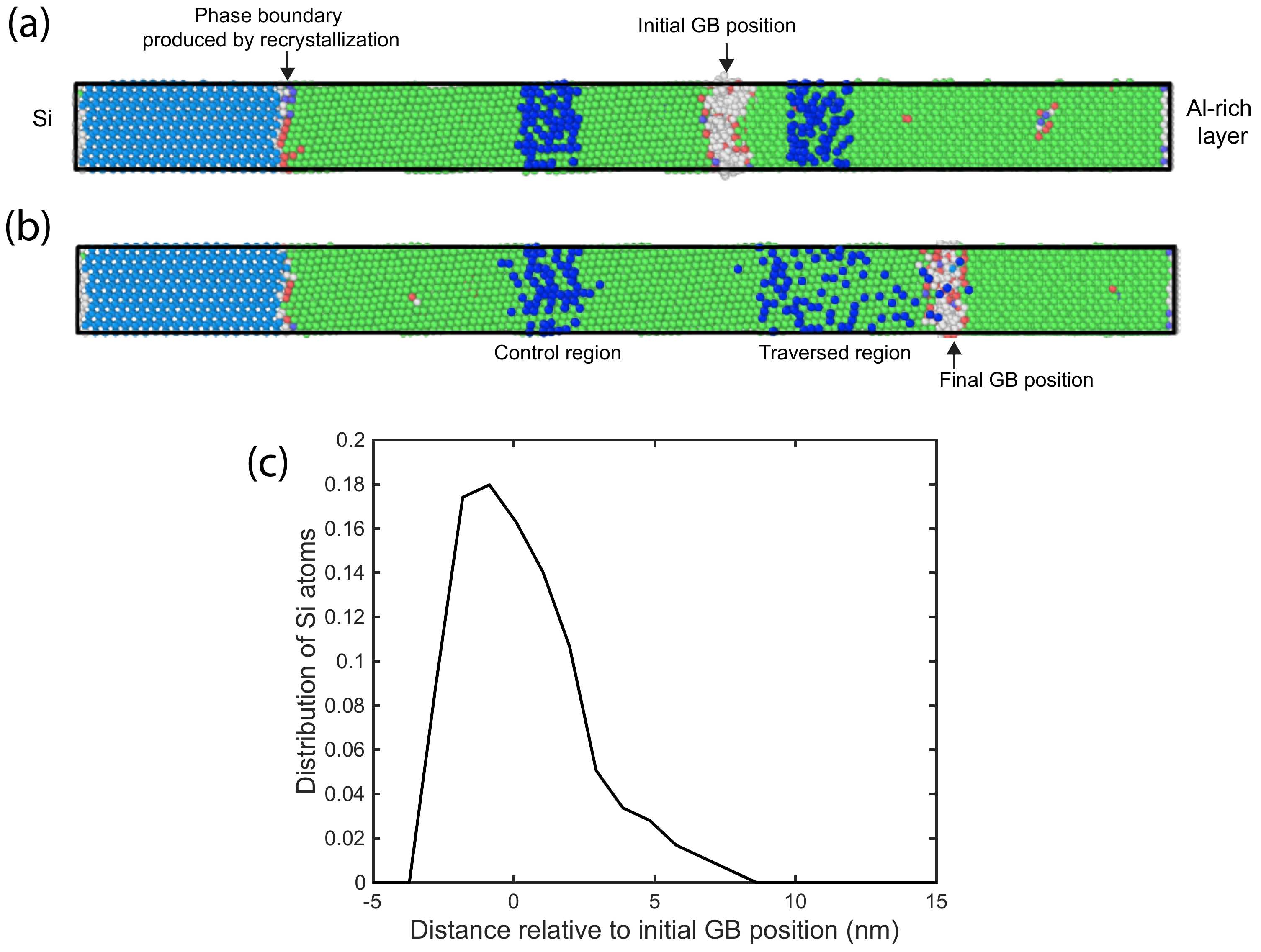}
\caption{Solute drag during GB motion caused by interface-induced recrystallization.
(a) Si atoms in two stripes behind and in front of the moving GB are
colored in blue for tracking. (b) The GB scatters the Si atoms in
the stripe traversed by the GB migration, providing evidence of Si
drag by the moving GB. (c) The solute drag is further confirmed by
the long tail in the Si composition profile after the right-hand stripe
was overrun by the GB.}
\label{fig:figS4}
\end{figure}

When the recrystallization triggers a significant GB migration, the
process is accompanied by a solute drag. In the example shown in Fig.~\ref{fig:figS4},
the crystallographic orientation across the interphase boundary is
the same as in Fig.~\ref{fig:figS3} but the interface cross-section
is relatively small and the lattice misfit is large (4\%), creating
a large strain energy differential across the GB. Fig.~\ref{fig:figS4}
demonstrates the drag of Si atoms by the moving GB using the technique
proposed in \citep{Koju:2020aa}. Si atoms were colored for tracking
purposes in two stripes parallel to the interface: one stripe was
chosen ahead of the moving GB and the other (control stripe) behind
the GB. As the GB traverses the right-hand stripe during the motion,
it scatters the Si atoms due to the accelerated (``short circuit'')
GB diffusion as well as by dragging the Si atoms in the direction
of motion. The diffusion-induced scattering alone would cause a nearly
even spreading of the Si composition profile in both directions. By
contrast, the solute drag spreads the solute atoms predominantly in
the direction of GB motion, creating a long tail in the composition
profile. This tail was clearly observed in our simulations (Fig.~\ref{fig:figS4}c),
providing evidence that the Si atoms were indeed dragged by the GB.

The previous direct observation of solute drag by MD simulations was
possible due to slow migration velocities ranging between 0.01 and
0.1 m/s \citep{Koju:2020aa}. Here, we observed the solute drag at
surprisingly large GB velocities of $\approx$ 1 m/s. The stronger
solute drag in the Al-Si system likely originates from the relatively
fast Si diffusion in Al and the highly disordered GB structure that
easily absorbs Si atoms.

\bigskip{}

\bigskip{}

\end{document}